\documentclass[useAMS,usenatbib]{mn2e}
\RequirePackage{xspace}
\usepackage{graphicx}
\usepackage{subfig}
\usepackage{tabularx}
\usepackage{amsmath}
\usepackage{color}
\usepackage{url}
\usepackage{amssymb}
\title[Instability of Magnetic Equilibria in Barotropic Stars] %% give here short title %%
{Instability of Magnetic Equilibria in Barotropic Stars}

\author[J.P. Mitchell et. al.]   %% give here short author list %%
{J.P. Mitchell$^1$ $^2$ , J. Braithwaite$^1$ , A. Reisenegger$^2$, H. Spruit$^3$, J.A. Valdivia$^4$, 
\newauthor and N. Langer$^1$ \\
$^1$Argelander-Institut, University of Bonn, Auf dem H\"{u}gel 71, 53121 Bonn, Germany \\ email: {\tt jmitchell@astro.uni-bonn.de}\\
$^2$Instituto de Astrof\'{i}sica, Facultad de F\'{i}sica, Pontificia Universidad Cat\'{o}lica de Chile, Av. Vicu\~na Mackenna 4860, 7820436 Macul, Santiago, Chile \\
$^3$Max-Planck-Institut f\"{u}r Astrophysik, Karl-Schwarzschild-Str. 1, D-85748 Garching, Germany\\
$^4$Departamento de F\'{i}sica, Facultad de Ciencias, Universidad de Chile, Casilla 653, Santiago, Chile}

\begin{document}

\date{Accepted . Received 1; in original form }

\pagerange{\pageref{firstpage}--\pageref{lastpage}} \pubyear{2014}

\maketitle

\label{firstpage}


\begin{abstract}
In stably stratified stars, numerical magneto-hydrodynamics
simulations have shown that arbitrary initial magnetic fields
evolve into stable equilibrium configurations, usually containing
nearly axisymmetric, linked poloidal and toroidal fields that
stabilize each other.  In this work, we test the hypothesis that stable stratification is a requirement for the existence of such stable equilibria. For this purpose, we follow numerically the evolution of magnetic fields in barotropic (and thus neutrally stable) stars, starting 
from two different types of initial
conditions, namely random disordered magnetic fields, as well as
linked poloidal-toroidal configurations resembling the previously
found equilibria. With many trials, we always find a decay of the
magnetic field over a few Alfv\'en times, never a stable
equilibrium.  This strongly suggests that there are
no stable equilibria in barotropic stars, thus clearly
invalidating the assumption of barotropic equations of state often
imposed on the search of magnetic equilibria. It also supports the
hypothesis that, as dissipative processes erode the stable
stratification, they might destabilize previously stable magnetic
field configurations, leading to their decay.

\end{abstract}
\section{Introduction}

There are several kinds of stars, namely upper main sequence
stars, white dwarfs, and neutron stars, that, contrary to the Sun,
have magnetic fields that are organized on large scales and
persist unchanged over long time scales. Much of their interior is
stably stratified, due to (a) gradients of entropy in white dwarfs and
radiative envelopes of main sequence stars; and (b) gradients of
composition (relative abundances of particle species) in the case
of neutron stars. Clearly, no dynamo action is taking place in
these stars, so their magnetic fields must be in stable
hydromagnetic equilibrium states in which the Lorentz force is
balanced by small perturbations to the otherwise spherical
pressure and gravitational forces (with solid shear stresses in
neutron star crusts possibly also playing a role).

It has long been known that purely toroidal (azimuthal) and purely
poloidal (meridional) magnetic field configurations are unstable
\citep{Tayler_1973,Markey_1973,Wright_1973}. However,
3-dimensional magneto-hydrodynamics (MHD) simulations of
self-gravitating balls of highly conducting gas
\citep{Braithwaite_2004,Braithwaite_2006}
have shown their magnetic field to evolve
naturally from an initially random configuration to a nearly
axisymmetric structure in which poloidal and toroidal components
of comparable amplitudes stabilize each other, persisting for many
Alfv\'en times. Further work has suggested that the stable
stratification of the fluid plays an important role in stabilizing
these structures \citep{Braithwaite_2009,Akgun_2013}, and
\citet{Reisenegger_2009} has conjectured that there are no stable
magnetic configurations in stars that are barotropic, since they are not
stably stratified.

On the other hand, attempting to construct plausible axisymmetric
magnetic equilibria for these stars, several authors
\citep{Tomimura_2005,Yoshida_2006,Haskell_2008,Akgun_2008,Kiuchi_2008,Ciolfi_2009,Lander_2009,Duez_2010,Pili_2014}
have imposed a barotropic equation of state, forcing the magnetic
field structure to satisfy the non-linear and thus highly
non-trivial Grad-Shafranov equation \citep{Grad_1958,Shafranov_1966}. This assumption is not only
very restrictive and unjustified for any of the stars likely to
contain hydromagnetic equilibria (only very massive,
radiation-dominated stars or extremely cold white dwarfs might
come close to being barotropic); it might even be incompatible
with having stable equilibria if the conjecture mentioned above is
correct.

On the other hand, stable stratification will in the long term be
eroded by non-ideal MHD processes such as heat diffusion (in main
sequence and white dwarf stars), beta decays, and ambipolar
diffusion (the latter two in neutron stars; see
\citealt{Reisenegger_2009}). If the above conjecture is true,
these processes would destabilize the magnetic field on these long
time scales, making it decay (unless stabilized, e.g., by the solid neutron star crust).

Thus, it is important to clarify whether stable magnetic
equilibria can exist in barotropic stars. A previous study by
\cite{Lander_2012}, based on a perturbative approach, gives a
negative answer. The present work explores the same question more
extensively, using full MHD simulations, in which we have evolved
initially disordered magnetic fields as well as the ordered,
axially symmetric, twisted-torus magnetic equilibria written
analytically by \cite{Akgun_2013}, in both barotropic and (for
comparison) stably stratified stars, in order to see whether
stable equilibria can be reached.

Section \ref{physical} gives a brief introduction of the effect that
stratification or its absence is expected to have on the stability of magnetic equilibria. 
Section \ref{modelsec} explains the numerical scheme
used for our simulations. In Section \ref{randomsec}, we explain
how the disordered fields are created, and show the results of
their evolution.  In Section \ref{equilsec}, we present the
axially symmetric equilibria we use, as well as their
evolution. The conclusions from our results can be found in
Section \ref{concsec}.

\section{Stable stratification vs. barotropy.}\label{physical}
The physical effect of stable stratification can be illustrated by
the thought experiment in which a fluid element is taken from a
given position and moved vertically against gravity to a new
position. Then, it is allowed to achieve pressure equilibrium
with its new surroundings, without allowing it to change its
entropy or chemical composition (which generally occur on much
longer time scales). If the entropy or chemical composition vary
with depth in the star, they will be different inside and outside
the displaced fluid element, causing its density to be different,
and thus creating a restoring force that is
quantified by the squared Brunt-V\"ais\"al\"a (or buoyancy)
frequency (force per unit mass per unit displacement)
\begin{equation}
\label{buoyancy} N^2=g^2\left[\left(\partial\rho\over\partial
P\right)_\mathrm{eq}-\left(\partial\rho\over\partial
P\right)_\mathrm{ad}\right].
\end{equation}
Here, $g$ is the local acceleration of gravity, $\rho$ is the
fluid mass density, $P$ is its pressure, and the subscripts ``eq''
and ``ad'' refer to the equilibrium profile existing in the star
and to the changes produced in the adiabatic displacement of the
fluid element, respectively. If $N^2>0$, the restoring force tends
to move the fluid element back to its initial position, causing the fluid
to be stably stratified, whereas $N^2<0$ causes a runaway
fluid element, resulting in a convective instability. If, on the
other hand, the fluid is barotropic, there is a one-to-one
relation between pressure and density, $P=P(\rho)$ (i.~e. because
the specific entropy and/or composition are uniform throughout the
fluid or adjust to an equilibrium as fast as the fluid element can
move), the two partial derivatives will be equal, and the fluid
will be neutrally stable.

In a hydromagnetic equilibrium state, the forces on a given fluid
element must be balanced,
\begin{equation}\label{balance}
-\nabla P-\rho\nabla\psi+{1\over c}\vec j\times\vec B=0,
\end{equation}
where $\psi$ is the effective gravitational potential (including a
centrifugal contribution in a rotating star), $\vec
j=c\nabla\times\vec B/(4\pi)$ is the electric current density, and
$\vec B$ is the magnetic field. From this, it follows that
\begin{equation}\label{eqcurl}
{\nabla P\times\nabla\rho\over\rho^2}=\nabla\times\left(\vec
j\times\vec B\over\rho c\right).
\end{equation}
In the barotropic case, $\nabla P$ and $\nabla\rho$ must be parallel
everywhere, so the left-hand side vanishes, forcing the right-hand
side to vanish as well, and thus imposing a strong constraint
(three scalar equations) on the magnetic field configuration. This
constraint is much weaker for the stably stratified case. In order
to understand the latter, we make the astrophysically well
justified assumption that the last term in Eq.
(\ref{balance}) is much smaller than the other two, so the
thermodynamic variables can be written as $P=P_0+P_1$ and
$\rho=\rho_0+\rho_1$, where $P_0$ and $\rho_0$ are their values in
an unmagnetized star, and $P_1$ and $\rho_1$ are small perturbations
caused by the Lorentz force, which can be regarded as
\emph{independent} because of the third, relevant thermodynamic
variable entering their relation (small perturbations of specific
entropy or chemical composition; see also \citealt{Reisenegger_2009,Mastrano_2011}). In this case, $\nabla
P\times\nabla\rho\approx\nabla
P_0\times\nabla\rho_1+\nabla P_1\times\nabla\rho_0$, so the
left-hand side of Eq. (\ref{eqcurl}) is generally non-zero,
only being constrained to be a horizontal vector field
(perpendicular to $\nabla P_0$). Thus, only the vertical component
of the curl on the right-hand side is required to vanish (a single
scalar equation), a much weaker constraint. The particular,
axially symmetric case is discussed in Section \ref{equilsec}.

In realistic stars, the characteristic Alfv\'en frequency
$\omega_A$ associated with magnetically induced motions (e.~g.,
the growth rate of magnetic instabilities) is much smaller than
$N$, thus the buoyancy will produce a powerful restoring force,
largely impeding magnetically induced vertical (radial) motions,
and thus strongly constraining possible magnetic instabilities.
If, on the other hand, we had $N=0$ (or $\ll\omega_A$), there
would be no such constraints on the motion, and the magnetic field
could decay much more easily. This general idea, as well as more
detailed arguments based on the same physics
\citep{Braithwaite_2009,Akgun_2013} motivate the conjecture that
there will be stable magnetic equilibria only in stably stratified
stars \citep{Reisenegger_2009}.

\section{The Models}\label{modelsec}
In this section we describe the setup of the model, starting with a
description of the numerical code.

\subsection{The numerical scheme.}
We use a three-dimensional Cartesian grid-based MHD code developed by
\cite{Nordlund_1995} (see also Gudiksen \& Nordlund 2005), which has
been used in many astrophysical contexts such as star formation, stellar
convection, sunspots, and the intergalactic medium (e.g. \cite{Padoan_2007,Braithwaite_2012,Collet_2011,Padoan_1999}). It has a staggered mesh,
so that different variables are defined at different locations in each
grid box, improving the conservation properties. The third-order
predictor-corrector time-stepping procedure of \cite{Hyman_1979} is used, and
interpolations and spatial derivatives are calculated to fifth and sixth
order respectively. The high order of the discretization is a bit more
expensive per grid point and time step, but the code can be run with fewer
grid points than lower order schemes, for the same accuracy. Because of the
steep dependence of computing cost on grid spacing (4th power for explicit
3D) this results in greater computing economy. We model the star as a ``star in a box'', by modeling a self-gravitating ball of gas inside of the cubic computational domain.  The simulations described
here, unless stated otherwise, are run at a resolution of $128^3$, and the star has a radius of $32$ grid spacings (see below).

For stability, high-order ``hyper-diffusive'' terms are employed for thermal,
kinetic, and magnetic diffusion. These are an effective way of
smoothing structure on small, badly-resolved scales close to the spatial
Nyquist frequency whilst preserving structure on larger length scales. This
results in a lower ``effective'' diffusivity compared to standard diffusion,
and is more computationally efficient than achieving the same by increasing
the resolution. In this study we are interested in physical processes
taking place on a dynamic timescale, and so the diffusion coefficients are
simply set to the lowest value needed to reliably prevent unpleasant
numerical effects (`zig-zags'), so that the diffusion timescale is as long
as possible. All three diffusivities are equal, i.e. the Prandtl and
magnetic Prandtl numbers are both set to $1$. The code contains a Poisson
solver to calculate the self-gravity.

\subsection{Timescales, numerical acceleration scheme}\label{times}
The presence of  several very different timescales  in stellar MHD problems presents computational difficulties. There is the sound crossing time $\tau_{\rm sound}$ (governing deviations from pressure equilibrium), the Alfv\'en crossing time $\tau_{\rm A}$ (on which a magnetic field evolves towards equilibrium),
%, the buoyancy time scale of the density stratification $N^{-1}$, 
the time scale $\tau_{\rm d}$ on which the magnetic field evolves under magnetic diffusion, and the rotation period of the star. We simplify the problem by assuming a non-rotating star, and by making sure the magnetic diffusion inherent in the numerical method is small enough that it does not significantly affect the evolution of the field, while still allowing for magnetic reconnection. The timescales to be included are then the sound travel time and the Alfv\'en time. For realistic Ap star field strengths of about a few kG, $\tau_{\rm A}$ is of order a few years, which is 5 orders of magnitude longer than the sound crossing time. To cover a range like this, we make use of the fact that  in a star close to hydrostatic equilibrium the evolution of the magnetic field of a given configuration depends on the field amplitude only through a factor in time scale. That is to say,  if a field ${\bf B}({\bf r},t)$ has the initial state ${\bf B}_0({\bf r})$, a field ${\bf B}^\prime$ with initial amplitude, ${\bf B}^\prime({\bf r},0)=k{\bf B}_0$, where $k$ is a constant, evolves approximately as
\begin{equation} {\bf B}^\prime({\bf r},t)=k{\bf B}({\bf r},kt). \label{scaling}\end{equation}
In other words, the time axis scales in proportion to the Alfv\'en crossing time. This approximation is 
%more accurate the larger the separation between $\tau_{\rm s}$ and $\tau_{\rm A}$. 
valid as long as the Alfv\'en crossing time is sufficiently long compared with the sound crossing time that the evolution takes place close to hydrostatic equilibrium, and at the same time sufficiently short compared to the magnetic diffusion time scale. 

We make use of this scaling to speed up the computation. The field strength of the configuration is multiplied by a time dependent factor $f(t)$, chosen so as to keep the Alfv\'en crossing time approximately constant between time steps. With Eq. (\ref{scaling}) the resulting (unphysical) field ${\bf B}_{\rm num}({\bf r},t_{\rm num}) $ is related to the physical field ${\bf B}$ by ${\bf B}({\bf r},t)=1/f~{\bf B}_{\rm num}({\bf r},t_{\rm num})$, where $t$ is related to $t_{\rm num}$ by ${\rm d}t=f{\rm d}t_{\rm num}$.  For the numerical implementation see the Appendix.
This {\em acceleration scheme} (Braithwaite \& Nordlund 2006) makes it possible to follow the decay of an unstable configuration to very low field strengths, by maintaining an artificially high amplitude magnetic field, and consequently shorter Alfv\'{e}n time, considerably decreasing computational needs and making such studies feasible.

Since the Alfv\'en speed is not uniform within the star, it is necessary to decide on some suitable average for the definition of $\tau_{\rm A}$; we use
\begin{equation}
\tau_{\rm A} = \frac{R}{\bar{v_{\rm A}}} = \frac{R\sqrt{4\pi\bar{\rho}}}{\bar{B}} = R\sqrt{\frac{M}{2E_{\rm mag}}},
\end{equation}
where $R$, $M$, $\bar{\rho}$, $\bar{B}$, and $E_{\rm mag}$ are the radius, mass, average density, average magnetic field, and total magnetic energy of the star. Similarly, the sound crossing time is defined as
\begin{equation}
\tau_{\rm{sound}}=R\sqrt{\frac{M}{\Gamma(\Gamma-1)E_{\rm th}}},
\end{equation}
where $E_{\rm th}$ is the total thermal energy content of the star and $\Gamma$ is the adiabatic index, further discussed in Section \ref{stablevsbarosec}.

The initial field strength of the configuration $\bar B_0$, or equivalently the value of the initial Alfv\'{e}n time $\tau_{\rm Ao}$, is an adjustable parameter. Its choice involves a compromise. A lower value of the initial field strength increases the separation between $\tau_{\rm{sound}}$ and $\tau_{\rm Ao}$. The approximation made is then better, but computationally more expensive. The value used here corresponds to  $\tau_{\rm{Ao}/}\tau_{\rm{sound}}\approx 15$.

\subsection{Implementing stably stratified vs. barotropic stellar models}\label{stablevsbarosec}
The code assumes a chemically uniform, monatomic, classical ideal
gas, whose specific entropy is $s\propto\ln(P/\rho^\Gamma)+ const.$, and
\begin{equation}
\label{adiabatic} \Gamma\equiv\left(\partial \ln
P\over\partial \ln\rho\right)_{\rm ad}={5\over 3}.
\end{equation}
This partial derivative as given corresponds to an adiabatic
change as discussed in \S~\ref{physical}.

Simulations so far (e.~g., \citealt{Braithwaite_2004,Braithwaite_2006,Braithwaite_2009}) have taken an initial density profile
inside the star corresponding to a polytrope
$P\propto\rho^\gamma$, with
\begin{equation}
\label{equilibrium} \gamma\equiv 1+{1\over n}\equiv\left(\partial
\ln P\over\partial \ln\rho\right)_\mathrm{eq}={4\over 3},
\end{equation}
where $n$ (here $=3$) is the usual ``polytropic index'', and the
value was chosen to roughly match the radiative zones of Ap stars.
For these values, since $\Gamma>\gamma$, we have $ds/dr>0$ and
$N^2>0$, so the star is stably stratified. We use this model as a
reference against which to compare the new barotropic model for
which we choose an initial profile with $\gamma=5/3$ ($n=3/2$), so
now $\Gamma=\gamma$, $ds/dr=0$, and $N^2=0$.

\begin{figure}
% \vspace*{-2.0 cm}
\begin{center}
(a){\includegraphics[width=0.475\textwidth]{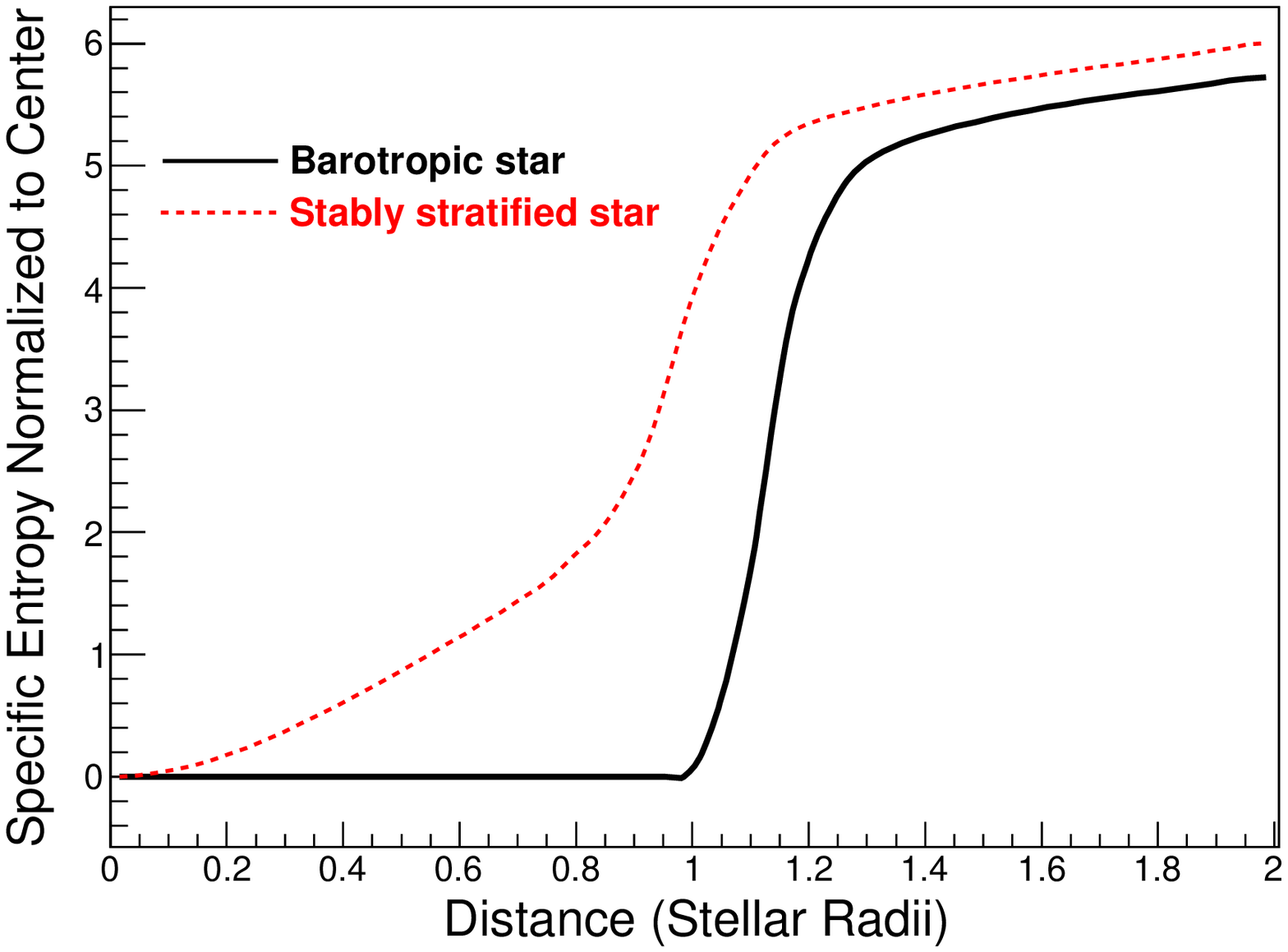}}\\
(b){\includegraphics[width=0.475\textwidth]{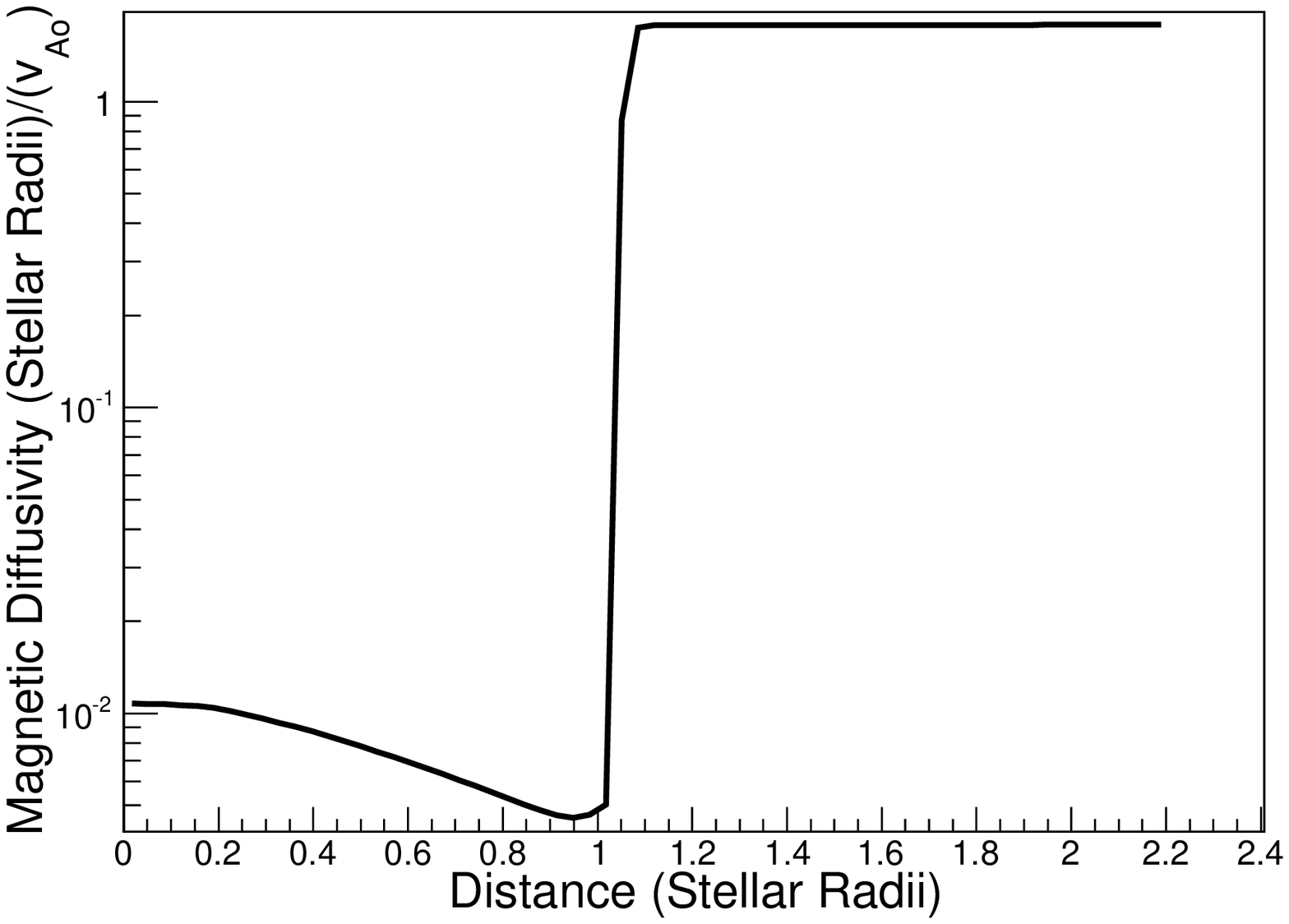}}
% \vspace*{-1.0 cm}
\caption{(a)Initial dimensionless specific entropy, with zero point
at its central value, as a function of $r$. The red dashed line is
for the stably stratified model ($n=3$), and the solid black line
is for the barotropic model ($n=3/2$).  (b) The magnetic diffusivity versus radius for the stably stratified model.} \label{entss}
\end{center}
\end{figure}

The profiles of the specific entropies of both models can be seen
in Figure \ref{entss}(a). The star in either model is surrounded by a
low-density, poorly conducting atmosphere, which causes the
magnetic field outside of the star to relax to a potential field.  Figure \ref{entss}(b) shows the magnetic diffusivity profile for the stably stratified model.
The atmosphere has a high temperature, so that its density does not
become too small towards the edges of the computational domain.
        
Note, however, that the numerical code contains heat diffusion.
Since the chosen profiles correspond to a radially decreasing
temperature, the diffusion will reduce the temperature gradient,
increasing the specific entropy gradient and thus making the fluid
increasingly stably stratified. In order to counteract this effect
and keep the fluid barotropic, we introduce an ad-hoc term in the
evolution equation for the internal energy per unit volume $e$, namely,
\begin{equation}
\label{eq1} \frac{{\rm d}e}{{\rm d}t}=...+\frac{\rho T
(s_0-s)}{\tau_{\rm s}},
\end{equation}
where $T$ is the local temperature, $s_0$ is the initial value of
the specific entropy inside the star, and $\tau_{\rm s}$ is a
timescale (to be chosen) at which this ``entropy term'' forces the
star back to its initially isentropic structure.

\section{Disordered initial fields}\label{randomsec}

The setup of the disordered initial field is done in the same way as in \cite{Braithwaite_2004} and \cite{Braithwaite_2006}. This consists in assigning random phases to locations in three-dimensional wavenumber space to wavenumbers from that which corresponds to the size of the star up to wavenumbers corresponding to about four grid spacings. The amplitudes are scaled as a function of wavenumber as $k^{-4}$:
\begin{equation}
 A_{\rm{k}}=(\rm{cos}(2\pi x_{\rm 1})+i\rm{sin}(2\pi x_{\rm 2}))k^{-4},
\end{equation}
where $A_{\rm k}$ is the amplitude for wavenumber $k$, and $x_{\rm 1}$ and $x_{\rm 2}$ are two randomly created phases. A reverse Fourier transformation is performed to obtain a scalar field.  Two more scalar fields are produced in the same way and these become the three components of a vector potential. The vector potential is then multiplied by $\rm{exp}(-\frac{r^2}{r_{\rm m}^2})$, where $r_{\rm m}$  was set equal to roughly a quarter of the radius of the star.  This was done to concentrate the field in the inner quarter of the star, so that the field dies off sufficiently quickly outside the star.

From this potential field, the magnetic field is then computed by taking the curl, and then its magnitude is scaled so as to obtain the desired total magnetic energy, which was $1/400$ times the thermal energy.

\begin{figure}
{\includegraphics[width=0.475\textwidth,height=0.55\textwidth]{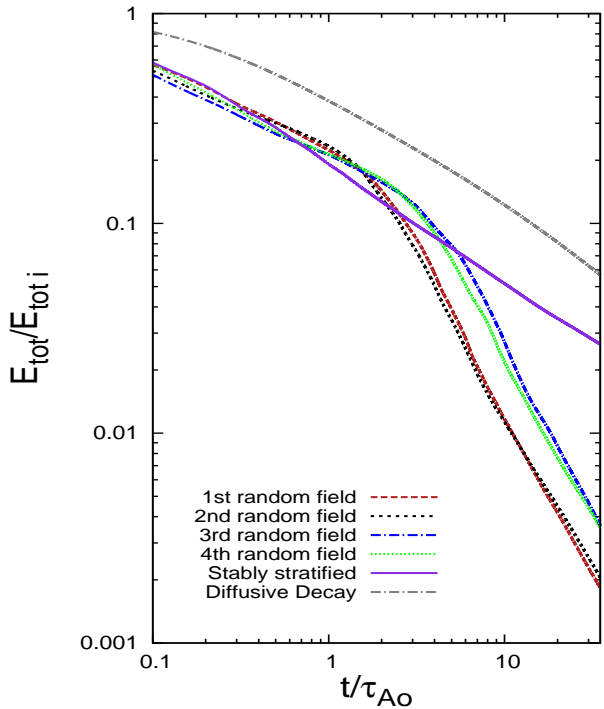}} \\
\caption{Total magnetic energy relative to initial total magnetic energy versus time, given in initial Alfv\'{e}n times, for models that have an initially disordered magnetic field configuration.  The solid violet curve is the evolution of a stably stratified model that reaches a stable equilibrium, and is shown for comparison.  The dashed-dotted gray line shows the evolution of the stably stratified model where the model is kept static, to show how the system will evolve under just diffusive processes.  All other curves are the evolution of barotropic models which began with initially different disordered fields, none of which reach stable equilibria.}  
   \label{fig1}
\end{figure}

\begin{figure}

 (a){\includegraphics[width=0.475\textwidth]{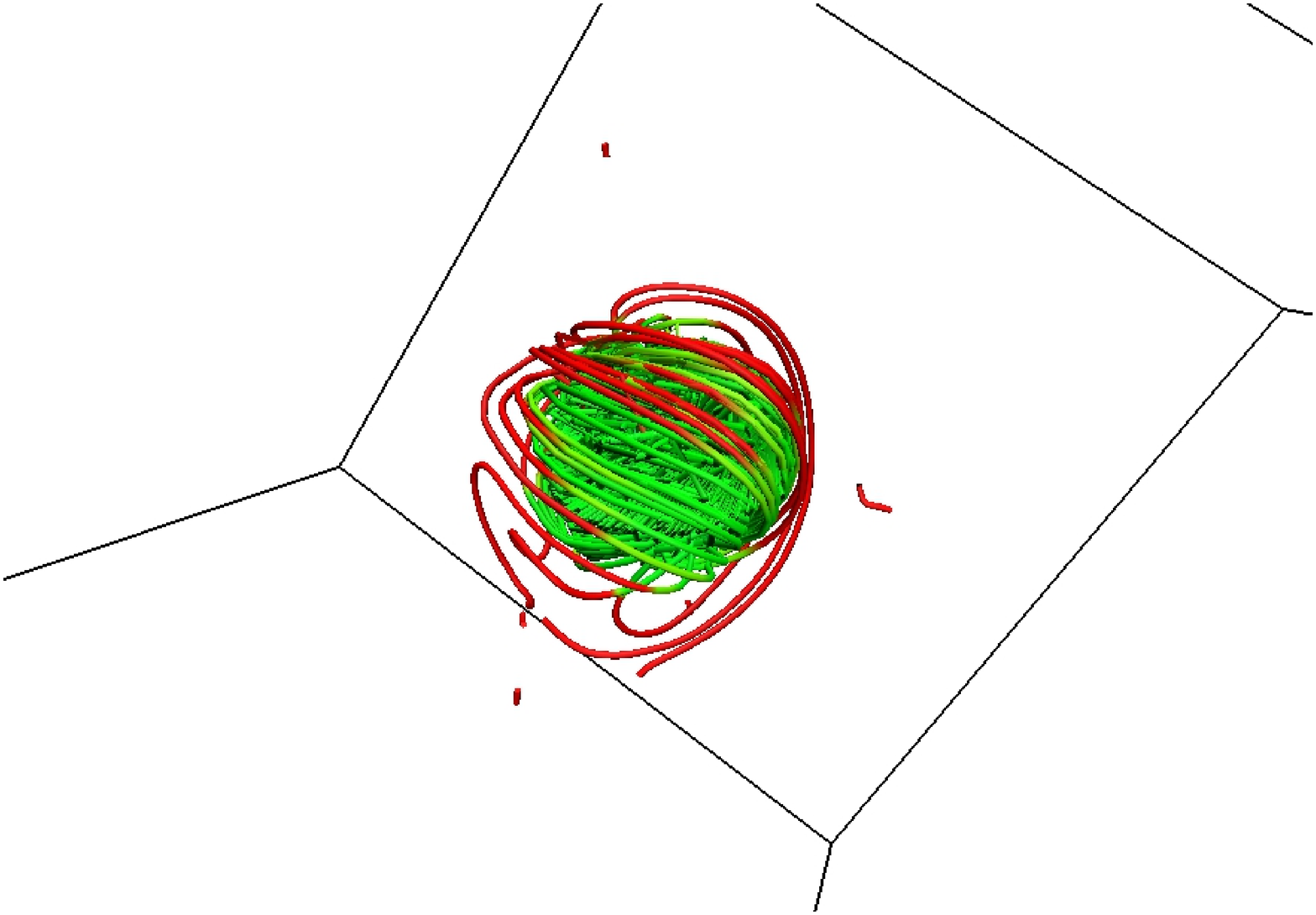}}
 (b){\includegraphics[width=0.475\textwidth]{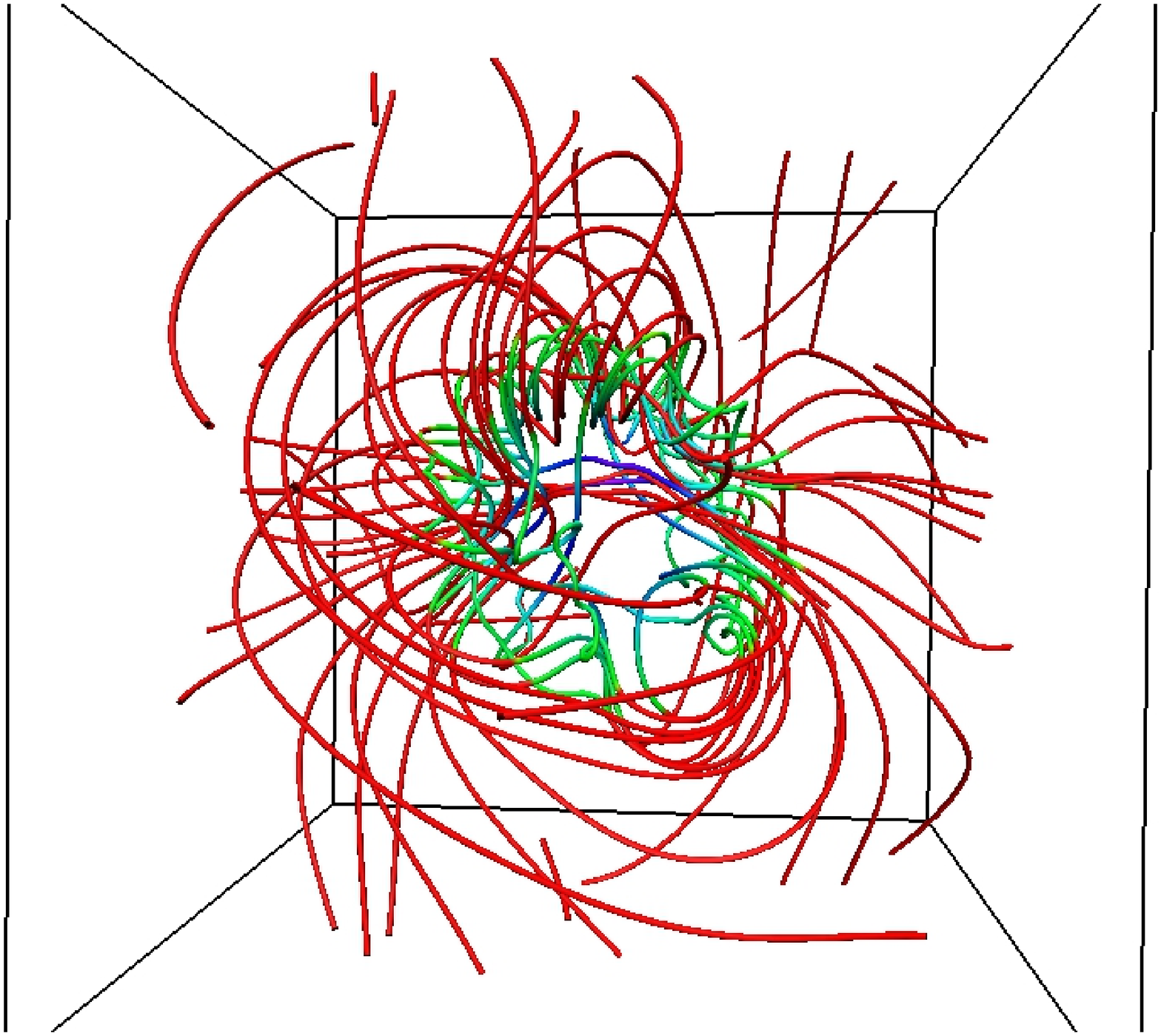}}
 \caption{Magnetic field configurations after a time of roughly 70$\tau_{\rm{Ao}}$ for models with initially disordered fields in a stably stratified star (panel (a)) and a barotropic star (panel (b)).  Lines shown in red are outside of the star, those that are blue and green are inside the star.  Panel (a) shows that the field has evolved into a large twisted torus inside the star.  It was shown that this configuration decays on a diffusive timescale and is stable.  Panel (b) shows that the barotropic model still has a disordered magnetic field, and much of the magnetic flux is at large radii or even outside of the star.}
   \label{randomeq}
\end{figure}

%\subsection{Disordered Initial Field Simulations}
%We begin with a set of models with an initially disordered magnetic field.  
We used four different barotropic
models, each including different initially disordered field configurations and an entropy timescale value roughly equal to the sound crossing time.  We also include one model where a disordered magnetic field was put into a stably stratified model 
as a comparison.  The works of \cite{Braithwaite_2004} and \cite{Braithwaite_2006} have shown that disordered fields in a stably stratified model %\textit{n}=3 polytropes 
can reach stable equilibria.  The evolution of the total magnetic energy of all of these models can be found in Figure \ref{fig1}.  The conclusion is that the stably stratified model reaches a stable equilibrium in a few Alfv\'{e}n timescales after which, the field decays only on the long diffusive timescale, which in a main-sequence star is of the order $10^{10}$ yr.
On the other hand, none of the barotropic models reach such a stable equilibrium.  Figure \ref{randomeq} contains 3-D renderings of the magnetic field for both a stably stratified % \textit{n}=3 
model and a barotropic model %\textit{n}=3/2 polytrope 
after a time of 70$\tau_{\rm{Ao}}$.  In the stably stratified model, %\textit{n}=3 model, 
there is a visible twisted torus structure, while the barotropic model %\textit{n}=3/2 model 
still has a disordered magnetic field configuration.  This suggests that stable stratification is a crucial ingredient for reaching a stable MHD equilibrium from a disordered field.

\begin{figure}
% \vspace*{-2.0 cm}
\begin{center}
 \includegraphics[width=0.475\textwidth]{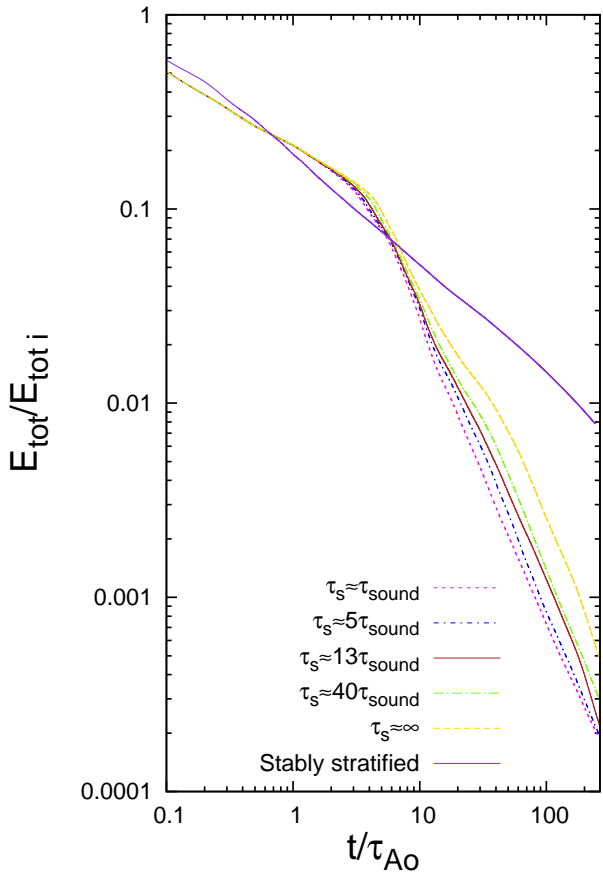}
%% \vspace*{-1.0 cm}
 \caption{Evolution of the total magnetic energy relative to its initial value, for models all starting from the same initially disordered magnetic field configuration, and evolved with values of $\tau_{\rm s}\approx$ 1, 5, 13, and 40 $\tau_{\rm sound}$, where $\tau_{\rm s}$ is the entropy timescale defined in Eq. \ref{eq1}, depicted as a short-dashed magenta, short-dashed dotted blue, solid red, and long-dashed dotted green curves, as well as a model in which the entropy term was not used, which is plotted as a long-dashed gold curve.  To compare, the evolution of the random magnetic field in a stably stratified model is also shown with the solid violet curve.  As the timescale is increased in the barotropic models, the magnetic energy decays more slowly at late times.  However, none of the simulations of barotropic models reaches a stable equilibrium.}
   \label{fig1b}
\end{center}
\end{figure}

We have also investigated the effect of using different values of $\tau_{\rm s}$ on the evolution of a disordered initial field.  Simulations were carried out with four different values of $\tau_{\rm s}$, as well as a simulation where the entropy term was not used, which is akin to $\tau_{\rm s}$ being equal to infinity.  Figure \ref{fig1b} shows the evolution of the total magnetic energy for these models, as well as the stably stratified one.  It is obvious that increasing the value of $\tau_{\rm s}$ has little effect on the evolution, although after a few $\tau_{\rm Ao}$, models with a larger entropy timescale decay slightly more slowly.  The reason for this is that the longer $\tau_{\rm s}$ timescale allows for a slight positive entropy gradient to evolve, thus a small buoyant restoring force will act to slow down the rise of the magnetic field.  It should be noted, however, that even in the case where the entropy term is not utilized in the code, the formation of a stable stratification does not occur quickly enough for a stable equilibrium to be created.

\section{Axisymmetric equilibria}\label{equilsec}

\cite{Akgun_2013} wrote down analytic expressions of linked poloidal-toroidal magnetic field configurations that correspond to hydromagnetic equilibria in stably stratified stars. (A particular form of these was also used by \cite{Mastrano_2011}) They have not been shown to be stable, and they are not equilibria in the barotropic case, but they might be stable equilibria (or close to these) in the stably stratified case and a good approximation to the best candidates for stability in the barotropic case.  

In order to motivate their functional form, consider a general, axisymmetric
field written as a sum of poloidal and toroidal components,
\begin{equation}
 \label{eqBfield}
  \boldsymbol{B}=\boldsymbol{B}_{\rm pol}+\boldsymbol{B}_{\rm tor}=B_{\rm p}\nabla \alpha(r,\theta)\times\nabla\phi +B_{\rm t}\beta(r,\theta)\nabla\phi,
\end{equation}
where $B_{\rm p}$ and $B_{\rm t}$ are characteristic values of the two components, and $\alpha$ and $\beta$ are dimensionless scalar functions
that depend on $r$, the radial coordinate, and $\theta$, the polar angle in
the star. Since there cannot be an azimuthal component of the
Lorentz force (which, in axial symmetry, cannot be balanced by
pressure gradients and gravity), $\nabla\alpha$ and $\nabla\beta$
must be parallel, thus one can write $\beta=\beta(\alpha)$.

The latter condition (a special case of the condition of the
vanishing vertical component of $\nabla\times(\vec j\times\vec B)/\rho c$
discussed in Section \ref{physical}) is the \emph{only} condition
required to be satisfied by $\alpha(r,\theta)$ and
$\beta(r,\theta)$ in order to yield a hydromagnetic equilibrium in
a stably stratified star. Besides this, these functions are
arbitrary and need not satisfy, e.~g., the commonly assumed
Grad-Shafranov equation, which is physically justified only for
barotropic fluids and otherwise represents an arbitrary,
additional constraint (e.~g., \citealt{Reisenegger_2009}). Of
course, in neither case there is a guarantee for the stability,
and thus for the astrophysical relevance, of the constructed
equilibria.

\begin{figure}
 \includegraphics[width=0.475\textwidth]{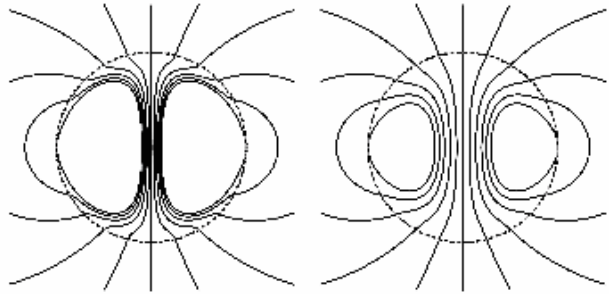}
 \caption{Cross sectional view of the poloidal field in the equilibria from \citet{Akgun_2013}, with the parameter $f_8$=-100 and -10 (left and right, respectively), controlling the volume of the closed field lines within the star.}
   \label{Akgun_model}
\end{figure}

For the poloidal part, \cite{Akgun_2013} choose
\begin{equation}
 \label{eqBpol}
\alpha(x,\theta)=f(x)\sin^{2}\theta
\end{equation}
where \textit{x} is the dimensionless radial coordinate, $x=r/R$, for $R$ the
stellar radius, and
\begin{equation}
 \label{eqBpolf8}
f(x)=\begin{cases} \sum_{i=1}^4 f_{2i}x^{2i} \mbox{     if $x \leq 1$} \\
      x^{-1} \mbox{     if $x > 1$,}
     \end{cases}
\end{equation}
where the $f_{2i}$ are constants. Outside the star, this gives an
exact dipole. For the internal field, three of the four constants
can be obtained from the condition that all components of the
magnetic field are continuous across the stellar boundary
(implying $f(1)=1$ and $f'(1)=-1$, where the primes denote
derivatives with respect to $x$) and that the current density
vanishes at the surface (implying $f''(1)=2$). The fourth constant, here chosen as $f_8$,
is a free parameter, which can be tuned to change the size of the
closed field line region of the equilibria, as seen in
Figure \ref{Akgun_model}.

The closed field line region is important because the toroidal
component of the field, described by the function $\beta(\alpha)$, can only exist in the volume
where the poloidal field lines are closed within the star. This motivates the choice:
\begin{equation}
 {\beta}= \begin{cases} ({\alpha}-1)^{\lambda} \mbox{     if ${\alpha}\geq$ 1} \\
  0  \mbox{     if ${\alpha} <$ 1,}
   \end{cases}
 \label{eqBtor}
\end{equation}
where the exponent $\lambda>1$, so that the current due to the
toroidal field decreases smoothly to zero at the stellar surface.

\subsection{Stability tests in stably stratified models.}\label{n3section}
We do not know a priori whether these equilibria will be stable in any type of star.  As a first step, we verified their stability in a stably stratified model.  If the magnetic field configurations are not stable in the stably stratified model, it seems likely that they will not be stable in a barotropic model.

We utilize the aforementioned equilibria of \cite{Akgun_2013} with the toroidal exponent $\lambda$ of Eq. \ref{eqBtor} set equal to 2, as was done in their analytic work, and $f_8$ from Eq. \ref{eqBpolf8} equal to -100.  The reason for choosing this value of $f_8$ was two-fold.  The equilibria found to evolve from random field configurations in the stably stratified models were found to have larger tori than the equilibria of \cite{Akgun_2013} in which the value of $f_8$ was set to zero.  Secondly, we chose $f_8$ to be a large amplitude, so as to be spread out the toroidal flux across a larger area of the star.  This would have the effect of decreasing the local magnetic energy density in the torus, as compared to a configuration with a smaller closed field line region with the same fraction of the toroidal to total magnetic energy.  In spreading out the toroidal energy, the local value of $\beta_{\rm plasma}$, defined as $\beta_{\rm plasma}=\frac{8{\pi}P}{B^{2}}$, in the torus would be larger, and it would thus be less buoyant.

\begin{figure}
% \vspace*{-2.0 cm}
\begin{center}
 \includegraphics[width=0.475\textwidth]{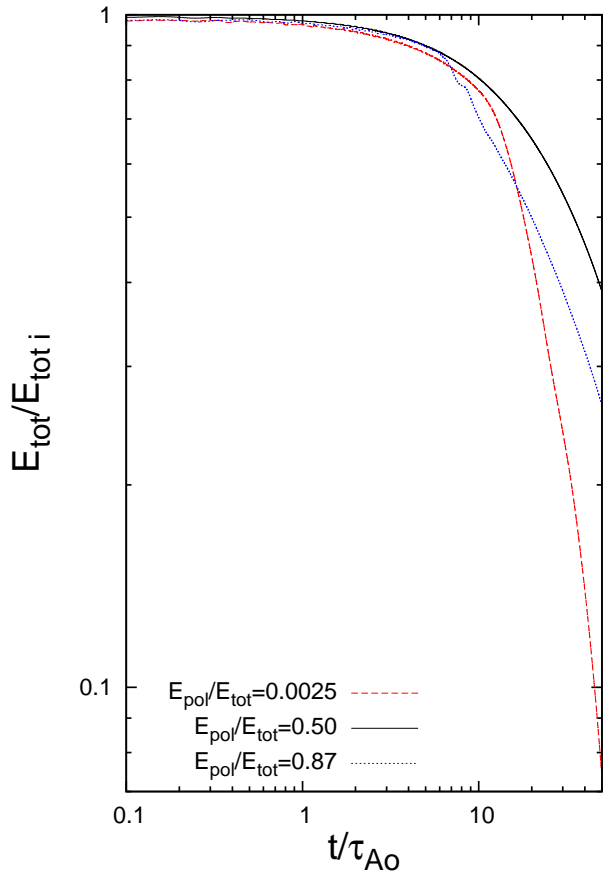}
% \vspace*{-1.0 cm}
 \caption{Evolution of the total magnetic energy given in initial Alfv\'{e}n times, for an initially ordered magnetic field with $\frac{E_{\rm pol}}{E_{\rm tot}}$ equal to 0.0025, 0.50, and 0.87 as the dashed red curve, solid black curve, and dotted blue curve respectively, in a stably stratified star.  }%Here, the model with an initial $\frac{E_{\rm pol}}{E_{\rm tot}}$ of 0.5 decays on a diffusive timescale, while the other models decay more quickly due to either \textit{m}=1 or \textit{m}=2 instabilities.}
   \label{n3_energyevolve}
\end{center}
\end{figure}

\begin{figure}
%\begin{tabular}{cc}
\subfloat{\includegraphics[width=0.2425\textwidth]{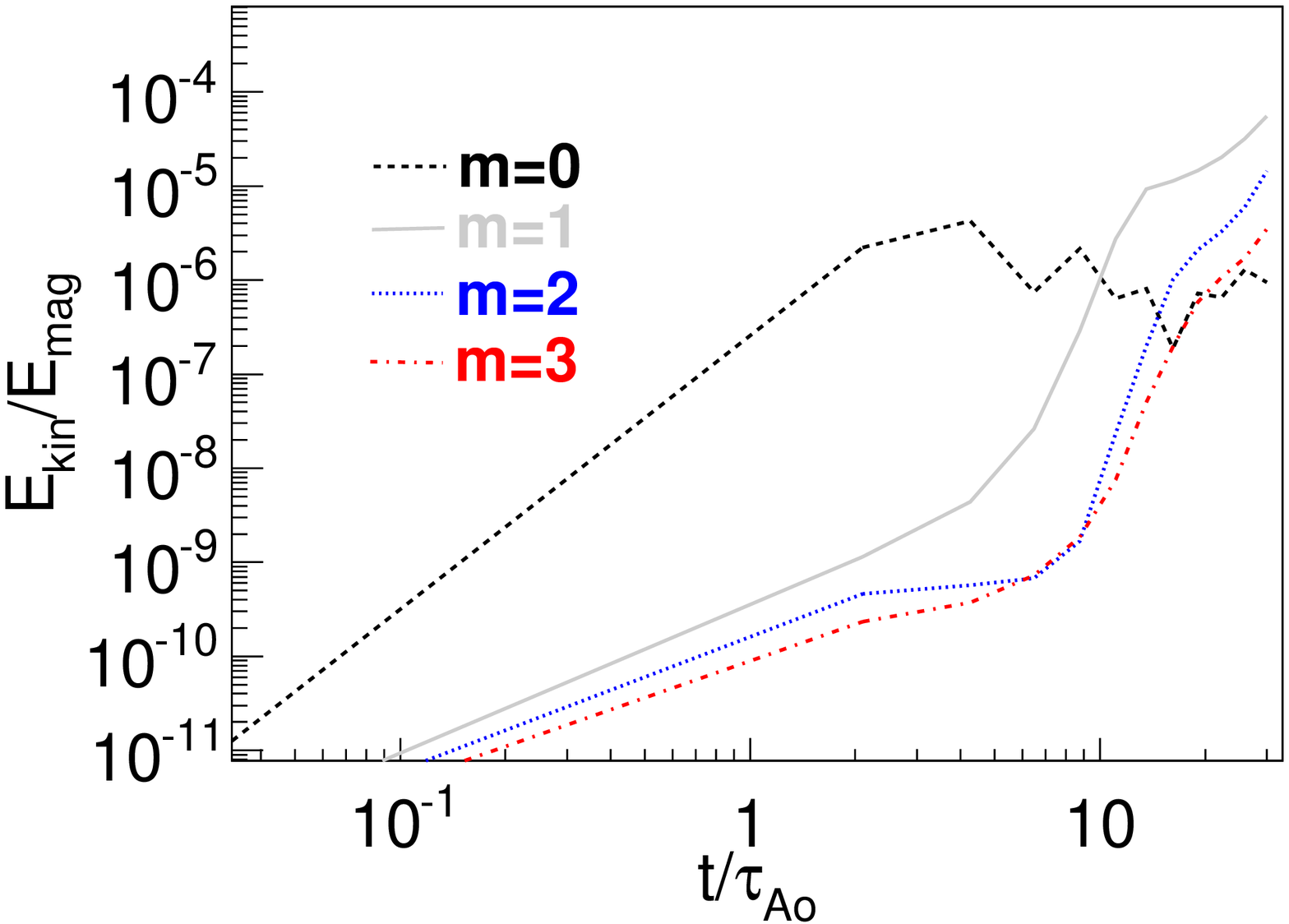}}
   \subfloat{\includegraphics[width=0.2425\textwidth]{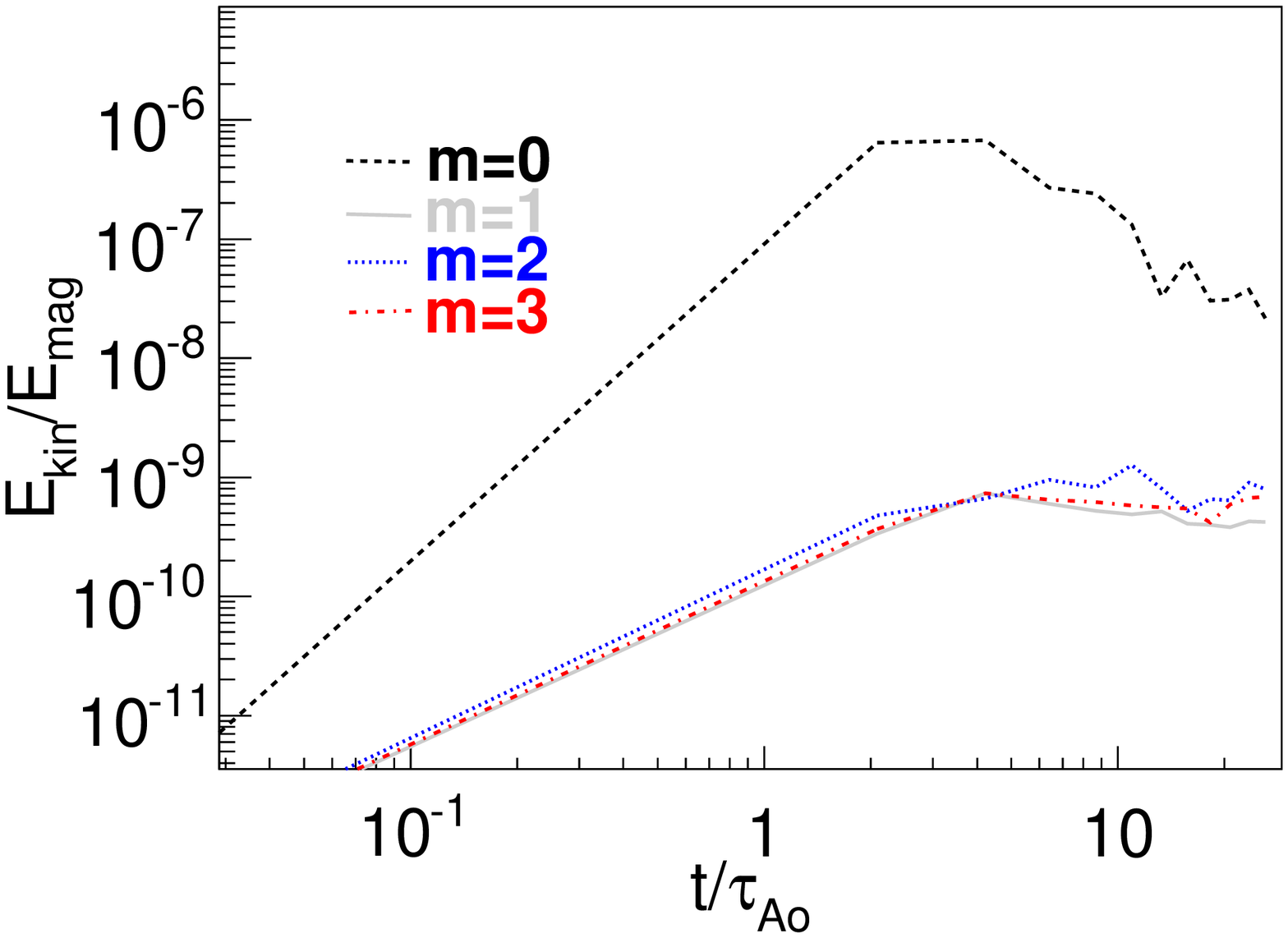}}\\
\subfloat{\includegraphics[width=0.2425\textwidth]{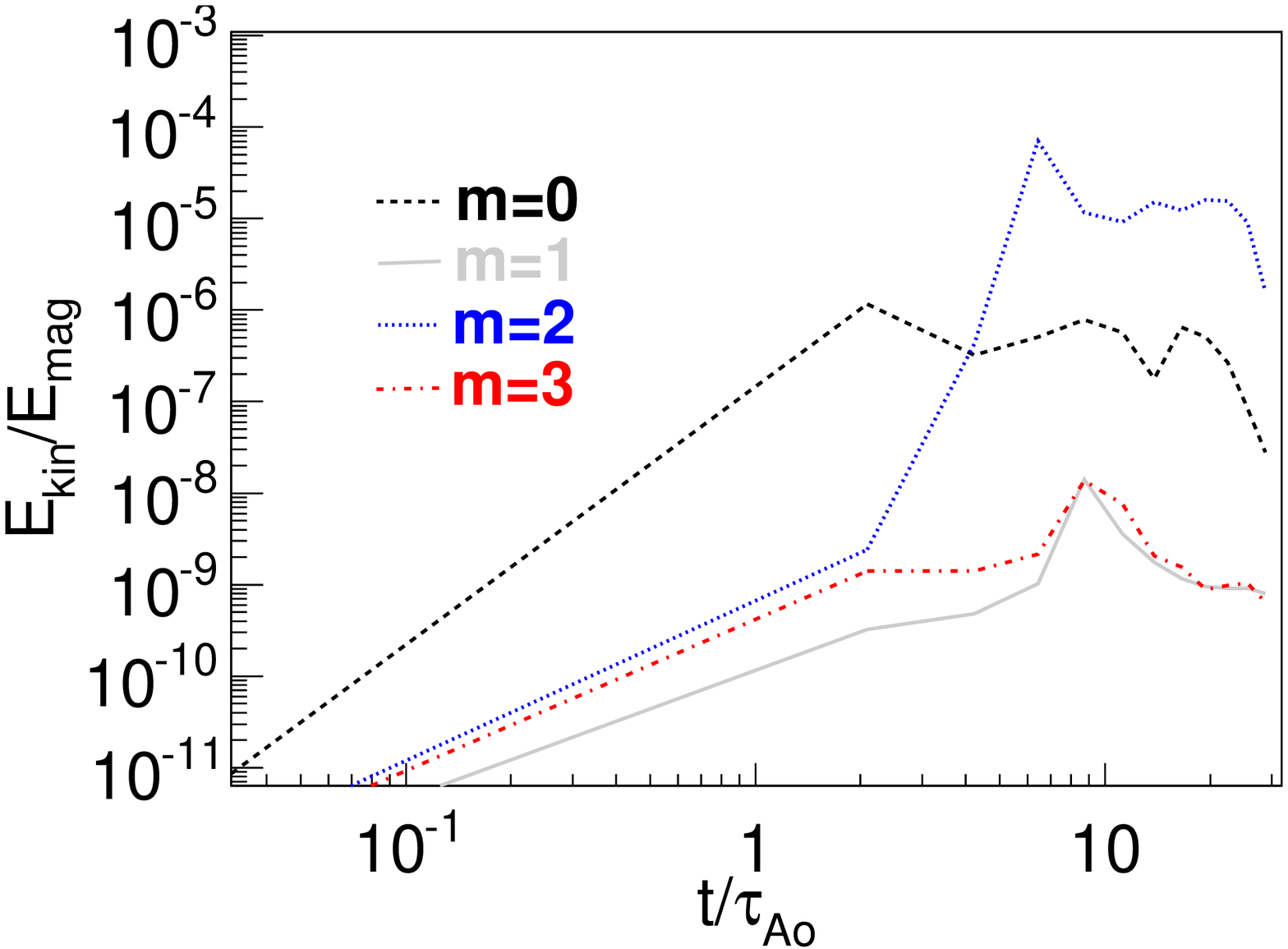}} \\
%\end{tabular}
\caption{$\theta$ component of the kinetic energy in each of the azimuthal \textit{m}=0-3 modes relative to the total initial magnetic energy versus time for stably stratified models with initially ordered magnetic fields and $\frac{E_{\rm pol}}{E_{\rm tot}}$ values of: 0.0025 (top left), 0.5 (top right), and 0.87 (bottom).  The dashed black curve represents the \textit{m}=0 mode, the solid gray curve the \textit{m}=1 mode, the dotted blue curve is the \textit{m}=2 mode, and the dashed-dotted red curve is the \textit{m}=3 mode.  }
\label{n3mmodes}
\end{figure}

We ran a number of simulations with various initial poloidal field energies relative to total magnetic energy $\frac{E_{\rm pol}}{E_{\rm tot}}$, ranging between 0.0015 and 0.93.  Such a large range was used, because at low values of this parameter, the Tayler instability \citep{Tayler_1973} is expected to set in, and at high values, instability along the ``neutral line'' may occur \citep{Markey_1973,Wright_1973}.  Models with initial $\frac{E_{\rm pol}}{E_{\rm tot}}$ between 0.008 and 0.8 were found to be stable, and to decay on a diffusive timescale, while all others decayed more quickly due to instabilities.  Figure \ref{n3_energyevolve} shows the evolution of the magnetic energy for some of the models.
To try to diagnose the source of instabilities, we analyzed the evolution of the $m$=0-3 azimuthal modes, which were determined by performing a root-mean-square integration of the $\theta$ component of the velocity in the meridional plane as:
\begin{equation}
  \begin{aligned}
&v_{\theta m}= \frac{1}{A}\\
&\int\sqrt{\left(\frac{1}{\pi}\int_0^{2\pi}v_{\theta}\cos(m\phi)\rm{d}\phi\right)^2+\left(\frac{1}{\pi}\int_0^{2\pi}v_{\theta}\sin(m\phi)\rm{d}\phi\right)^2}\rm{d}A,
    \end{aligned}
\end{equation}
where $v_{\theta m}$ is the amplitude of the theta component of the velocity for a particular $m$-mode, $A$ is the area of the star in the meridional plane, and $v_{\theta}$ is the $\theta$ component of the velocity.  We then calculated the $\theta$ component of the kinetic energy in each of these modes, and plotted these values normalized to the initial total magnetic energy versus time in 
Figure \ref{n3mmodes}.  % for models with initial $\frac{E_{\rm pol}}{E_{\rm tot}}$ values equal to 0.0025, 0.5, and 0.87.  
Here, it is evident that the model with $\frac{E_{\rm pol}}{E_{\rm tot}}$ of 0.5 is stable to all of the modes, as the curves for each mode evolve to a steady value.  The model with initial $\frac{E_{\rm pol}}{E_{\rm tot}}$ of 0.87 was unstable to \textit{m}=2 which can be seen by the sharp peak in the \textit{m}=2 curve at a time of roughly 2-5$\tau_{\rm{Ao}}$ after which this model calms to a new equilibrium as the \textit{m}=2 mode flattens out.  The model with initial $\frac{E_{\rm pol}}{E_{\rm tot}}$ of 0.0025 was unstable to  the \textit{m}=1 mode which begins to set in at a time of roughly 4$\tau_{\rm{Ao}}$, as can be seen by the peak in the figure.  It was then found that all configurations with initial $\frac{E_{\rm pol}}{E_{\rm tot}}$ above 0.8 were unstable to the \textit{m}=2 mode, while those with initial $\frac{E_{\rm pol}}{E_{\rm tot}}$ below 0.008 were unstable to the \textit{m}=1 mode.
In the \textit{m}=1 instability, the toroidal component of the field can be thought of as a stack of rings on top of one another; as the instability occurs, the rings slip with respect to each other.  In the \textit{m}=2 instability, the region along the neutral line, which is the space where the poloidal field goes to zero, experiences tension, which results in kinking of the neutral line, stretching any toroidal field that may be present along the neutral line.  This kinking of the neutral line bends the toroidal field lines into a shape that is similar to the seam of a tennis ball.  Both the $m$=1 and $m$=2 unstable configurations end up reaching another stable equilibrium after some time.  In the case of the $m$=1 unstable model, the equilibrium is reached at a much later time.  However, it can be seen that the $m$=2 unstable model reaches a new ``tennis ball'' shaped equilibria after roughly 8$\tau_{\rm{Ao}}$ when it begins to decay on a diffusive timescale, and the kinetic energy amplitude in the $m$=2 mode drops off and becomes stable in Figure \ref{n3mmodes}.  These results are similar to those of \cite{Braithwaite_2009}, who studied the stability of twisted torus configurations in stably stratified stars.

\subsection{Stability tests in barotropic models}
With the knowledge that these equilibria are stable in a stably stratified star for initial $\frac{E_{\rm pol}}{E_{\rm tot}}$ values between about 0.008 and 0.8,
we now investigate whether such stable equilibria can exist in barotropic stars.

\subsubsection{Existence of stable equilibria?}\label{n32section}
A series of models have been run with the same equilibria used in barotropic models.  For these models, $\tau_{s}$ was set to be equal to the sound crossing time of the star.  The initial $\frac{E_{\rm pol}}{E_{\rm tot}}$ values used were between 0.03 and 0.96.  The motivation for using values of $\frac{E_{\rm pol}}{E_{\rm tot}}$ above 0.9 comes from the works by \cite{Tomimura_2005,Ciolfi_2009,Lander_2009,Armaza_2013}, where magnetic equilibria have been calculated in axial symmetry for barotropic stars, and none of which obtained $\frac{E_{\rm pol}}{E_{\rm tot}}$ $<$ 0.9.  The rest of the spectrum of $\frac{E_{\rm pol}}{E_{\rm tot}}$ values was chosen based on the results of Section \ref{n3section}, where for a stably stratified star, the equilibria were stable for initial $\frac{E_{\rm pol}}{E_{\rm tot}}$ values of 0.008 to roughly 0.8.  In addition, \cite{Ciolfi_2013} have calculated magnetic equilibria in barotropic stars and found equilibria with $\frac{E_{\rm pol}}{E_{\rm tot}}$ values as low as roughly 0.11.

\begin{figure}
% \vspace*{-2.0 cm}
\begin{center}
 \includegraphics[width=0.475\textwidth]{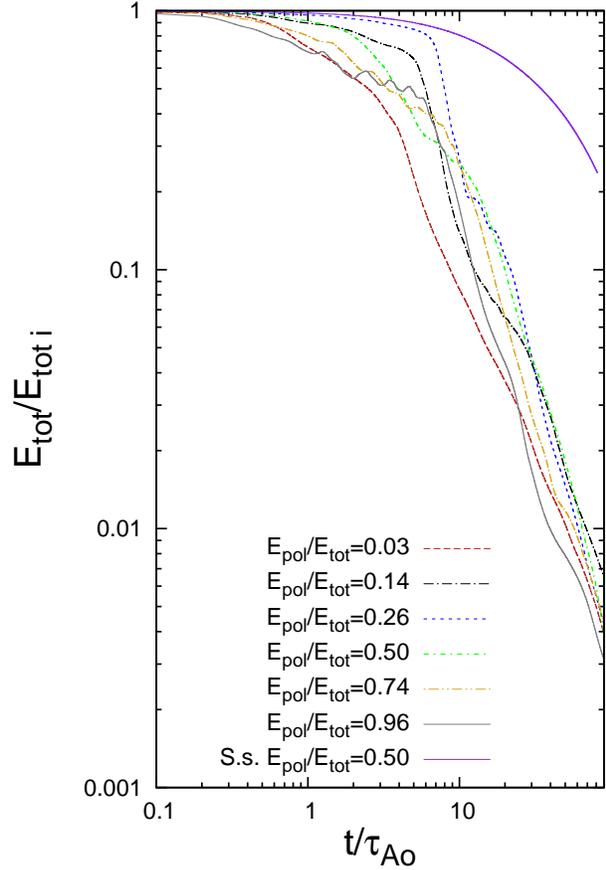}
 \caption{Evolution of the total magnetic energy relative to its initial value for a stably stratified star with an initially ordered field with $\frac{E_{\rm pol}}{E_{\rm tot}}=$0.5 as the solid violet curve, and a series of curves for barotropic stars with initial $\frac{E_{\rm pol}}{E_{\rm tot}}$ values of: 0.03 (red dashed), 0.14 (black dashed-dotted), 0.26 (blue short-dashed), 0.5 (green short-dashed dotted ), 0.74 (gold dashed two-dotted), and 0.96 (gray solid).  The stably stratified star's magnetic field decays on a diffusive timescale (see Section \ref{n3section}), while all of the barotropic stars' magnetic fields decay faster.}
   \label{n32Etotevolutionwithtext}
\end{center}
\end{figure}

\begin{figure}
% \vspace*{-2.0 cm}
\begin{center}
 \includegraphics[width=0.475\textwidth]{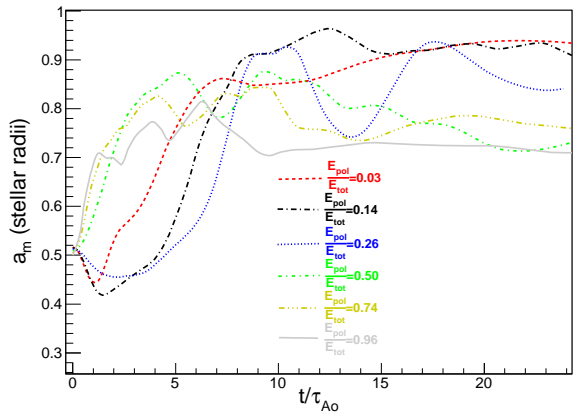}
 \caption{Evolution of $a_{\rm m}$, the effective magnetic radius (Eq. \ref{eqam}), for the barotropic models discussed in Figure \ref{n32Etotevolutionwithtext}. All models with an initial $\frac{E_{\rm pol}}{E_{\rm tot}}$ value less than about 0.33 experience an initial decrease in $a_{\rm m}$ and then subsequent increase as the torus initially contracts due to tension and then rises outwards.  All models with an initial $\frac{E_{\rm pol}}{E_{\rm tot}}$ above 0.33 simply experience an increase in $a_{\rm m}$ as the tori simply rise out of the star.  In all models the final state is the same, the magnetic field rises out of the star.}
   \label{n32am_and_epol}
\end{center}
\end{figure}

Figure \ref{n32Etotevolutionwithtext} shows the evolution of the total magnetic energy for barotropic models, as well as a stably stratified model with $\frac{E_{\rm pol}}{E_{\rm tot}}=$ 0.5 for comparison.  It is evident from the figure that none of the equilibria are stable in the barotropic models. An interesting point, however, is that fields with a initial $\frac{E_{\rm pol}}{E_{\rm tot}}$ values between 0.33 and 0.2 decay much more slowly at earlier times than all other barotropic models.  The reason for the slower decay of the magnetic field for these models can be seen in Figure \ref{n32am_and_epol}, where the effective magnetic radius, $a_{\rm m}$, defined as:
\begin{equation}
  a_{\rm m}^{2}=\frac{\int B^2r^2 \rm{d}V}{\int B^2 \rm{d}V},
  \label{eqam}
\end{equation}
 is plotted.  It is evident that for magnetic configurations with an initial $\frac{E_{\rm pol}}{E_{\rm tot}}\geq$0.33, $a_{\rm m}$ increases, caused by the torus rising in the star.  For  $\frac{E_{\rm pol}}{E_{\rm tot}}$ values below 0.33, the magnetic tension in the torus results in its contraction, causing a decrease of $a_{\rm m}$.  However, these models that initially experience a contraction of the torus still show that the $a_{\rm m}$ value will increase after roughly 6 $\tau_{\rm Ao}$, as the torus rises out of the star.

\subsubsection{Effect of $\tau_{s}$}

\begin{figure}
%% \vspace*{-2.0 cm}
\begin{center}
 \includegraphics[width=0.475\textwidth]{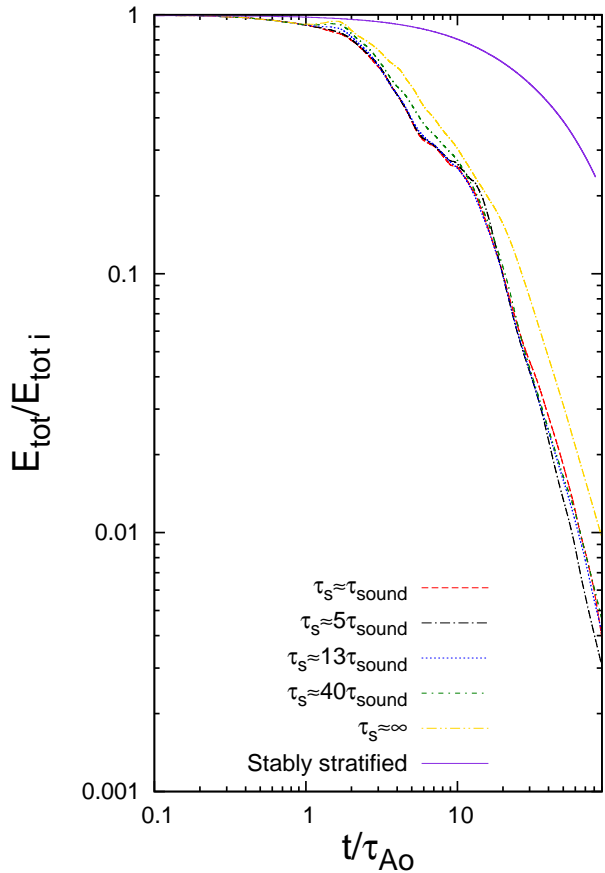}
 \caption{Evolution of the total magnetic energy relative to its initial value for models of barotropic stars with an initial $\frac{E_{\rm pol}}{E_{\rm tot}}$ value of 0.5, each with a different $\tau_{s}$ value.  $\tau_{s}\approx$ 1, 5, 13, and 40 $\tau_{\rm sound}$ are plotted as the dashed red, long-dashed dotted black, dotted blue, and short-dashed dotted green curves respectively.  In addition, one model where the entropy forcing term was not included is plotted as a gold dashed double-dotted curve, and the comparison model of a stably stratified star is plotted as the solid violet curve.  Regardless of the value of $\tau_{s}$ used in the barotropic models, the configuration is always unstable.}
   \label{tauvary}
\end{center}
\end{figure}

We also investigated the effect of varying $\tau_{s}$ for a magnetic configuration with an initial $\frac{E_{\rm pol}}{E_{\rm tot}}$ value of 0.5 in Figure \ref{tauvary}.  It is evident that a longer $\tau_{s}$ allows the magnetic field to decay more slowly.  As the entropy timescale becomes larger, the star, due to numerical diffusion, will evolve an entropy gradient, which will impede the rise of the torus.  However, even without the entropy adjustment term, the equilibria are still not stable.

\subsubsection{Variations of the free parameters in the axially symmetric equilibria.}

It was shown in Section \ref{n32section} that, regardless of the initial $\frac{E_{\rm pol}}{E_{\rm tot}}$ fraction used, all the magnetic configurations were unstable.  However, the initial fraction of the total magnetic energy in the poloidal component is just one of the free parameters that can be investigated in the axisymmetric configurations.  The effect of varying other free parameters, namely the value of $\lambda$ in Eq. \ref{eqBtor}, which affects the distribution of the toroidal component, the value of $f_{8}$ in Eq. \ref{eqBpolf8}, which controls the size of the closed field line region, and the radius of the ``neutral line'' were also studied.  We found that variations to these parameters had no effect on the final state of the simulations, as all configurations were found to be unstable to the same effect seen in Section \ref{n32section}, where the torus flowed radially out of the star, subsequently decaying in the atmosphere.

\subsection{Instabilities}
\begin{figure}
%\begin{tabular}{cc}
\subfloat{\includegraphics[width=0.2425\textwidth]{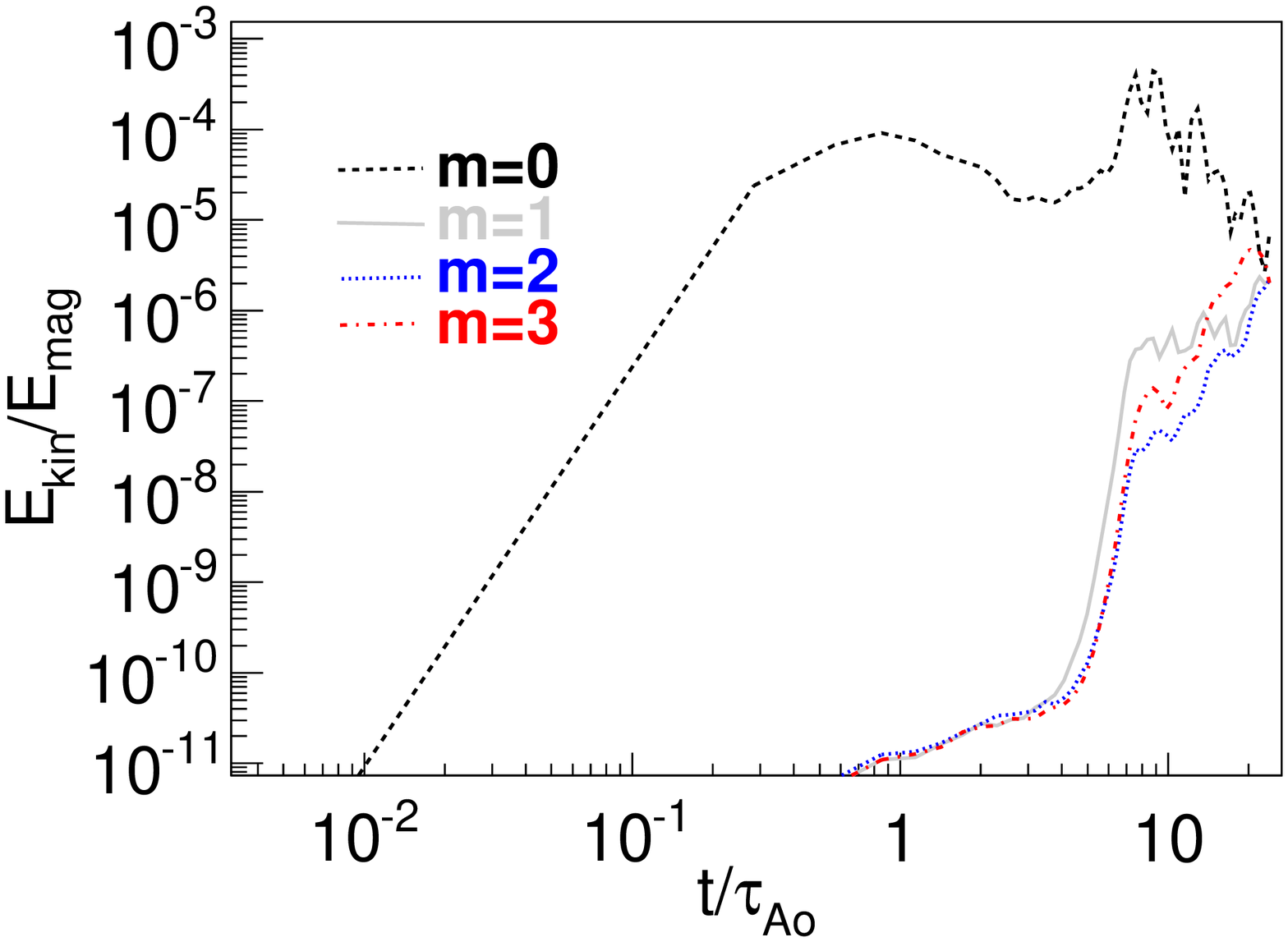}}
   \subfloat{\includegraphics[width=0.2425\textwidth]{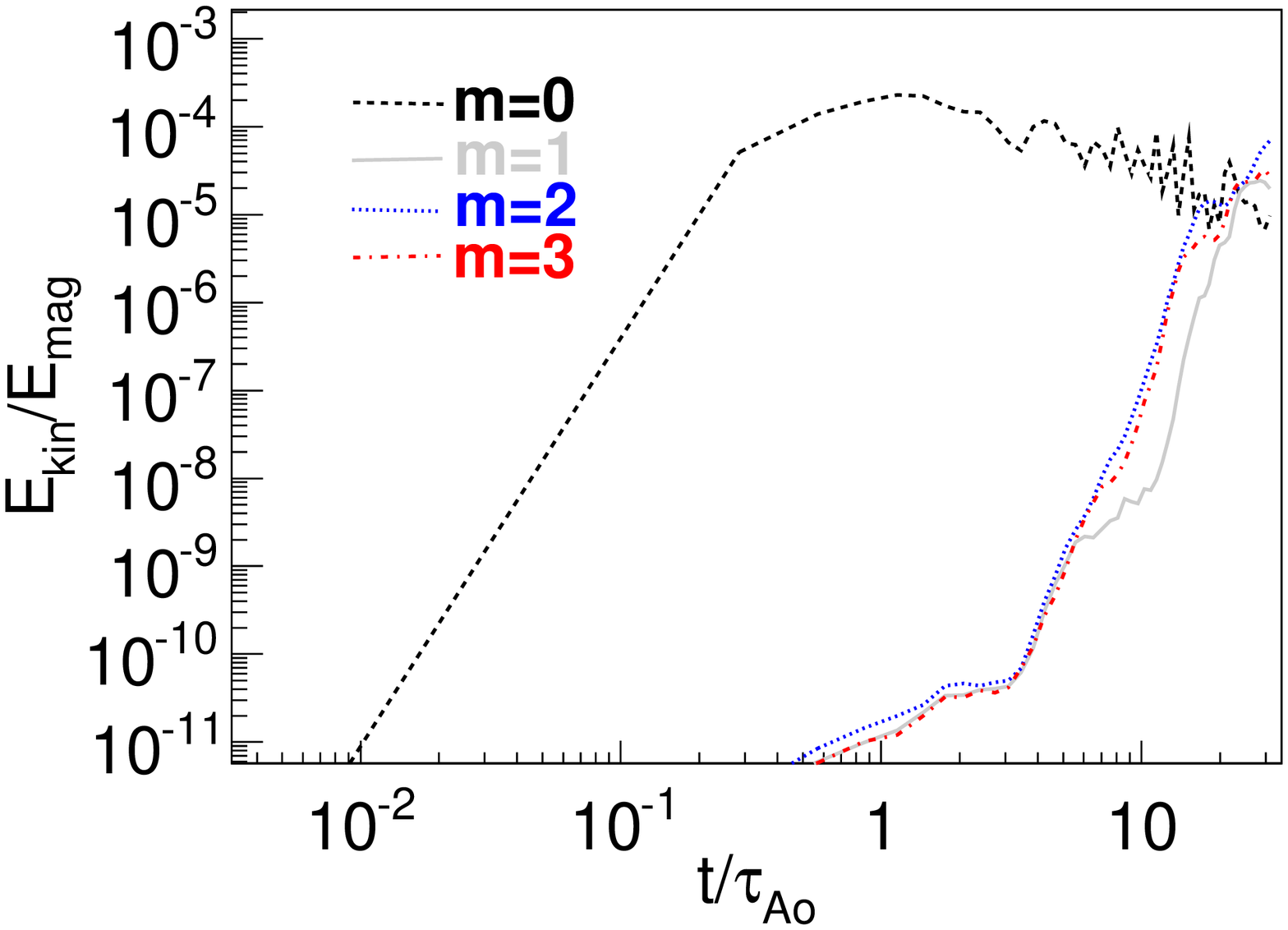}}\\
\subfloat{\includegraphics[width=0.2425\textwidth]{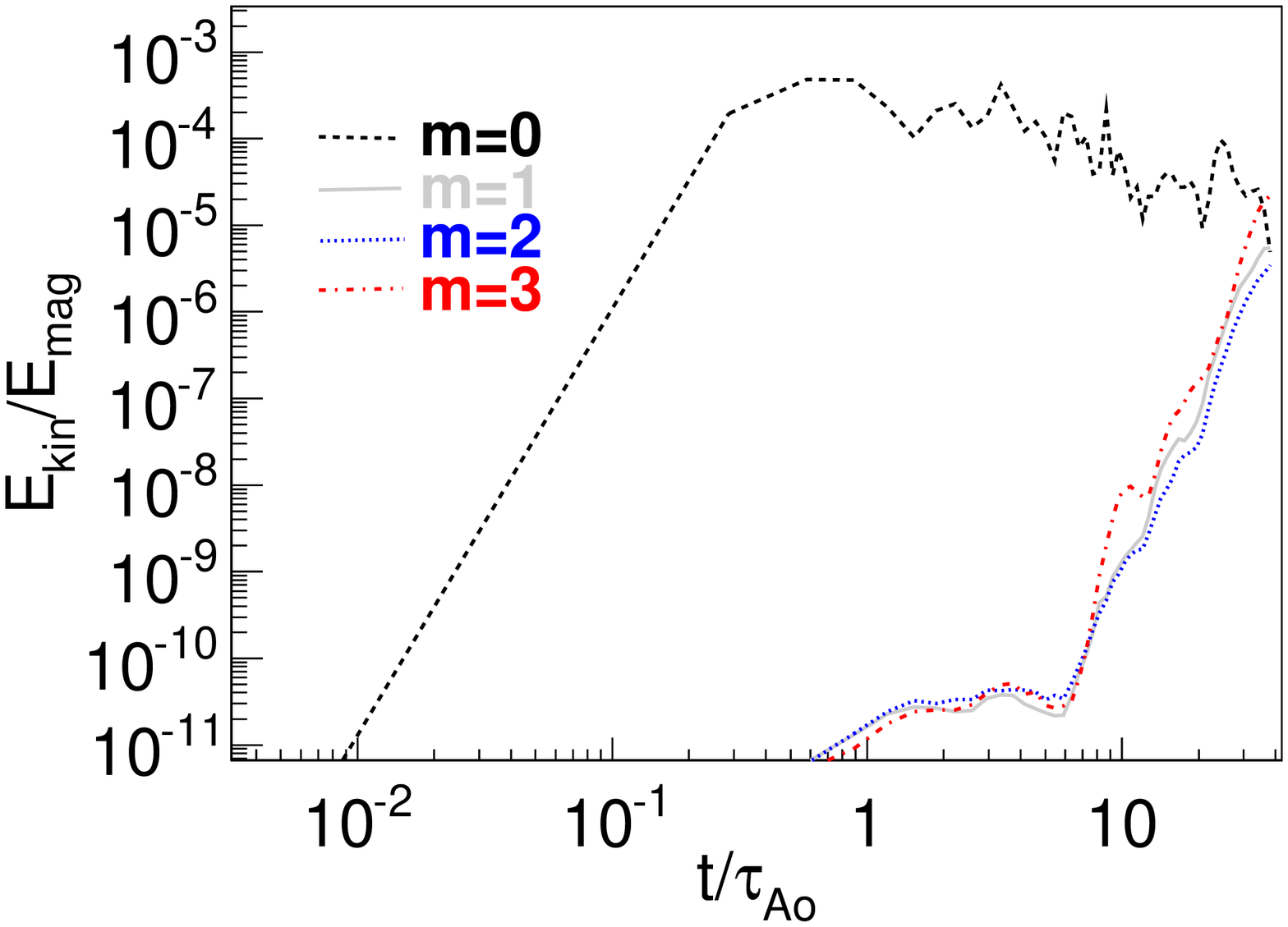}}
    \subfloat{\includegraphics[width=0.2425\textwidth]{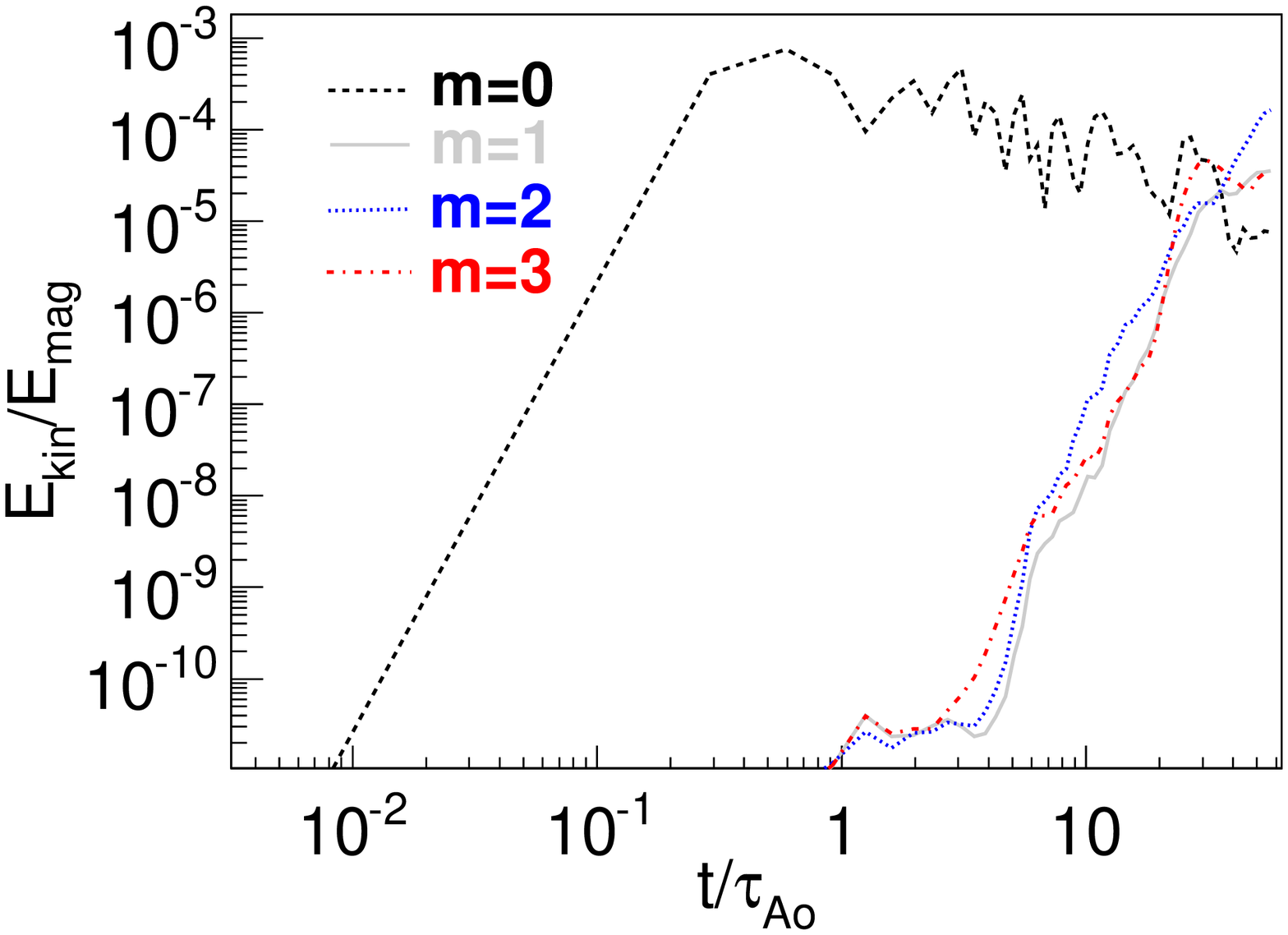}}\\
%\end{tabular}
\caption{Evolution of the $\theta$ component of the kinetic energy in each of the azimuthal \textit{m}=0-3 modes relative to the total initial magnetic energy, for barotropic stars containing initial magnetic field configurations with $\frac{E_{\rm pol}}{E_{\rm tot}}$ values equal to: 0.26 (top left), 0.5 (top right), 0.76 (bottom left), and 0.96 (bottom right).  In all cases the \textit{m}=0 mode is the dominant mode of instability.}\label{mmodes}
\end{figure}

\begin{figure}
\begin{tabular}{cc}
\subfloat{\includegraphics[width=0.15\textwidth,height=0.15\textwidth]{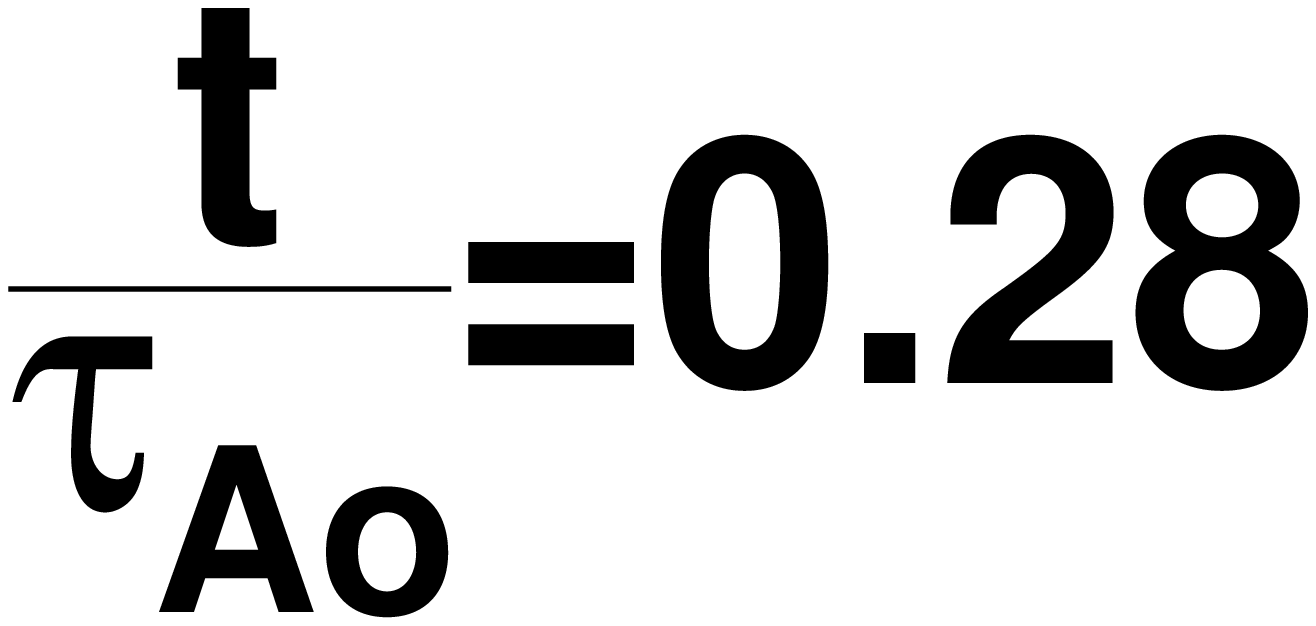}}
\subfloat{\includegraphics[height=0.225\textwidth]{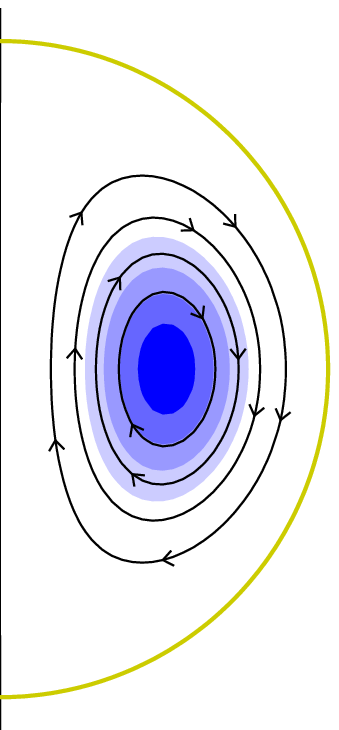}}\\
\subfloat{\includegraphics[width=0.15\textwidth,height=0.15\textwidth]{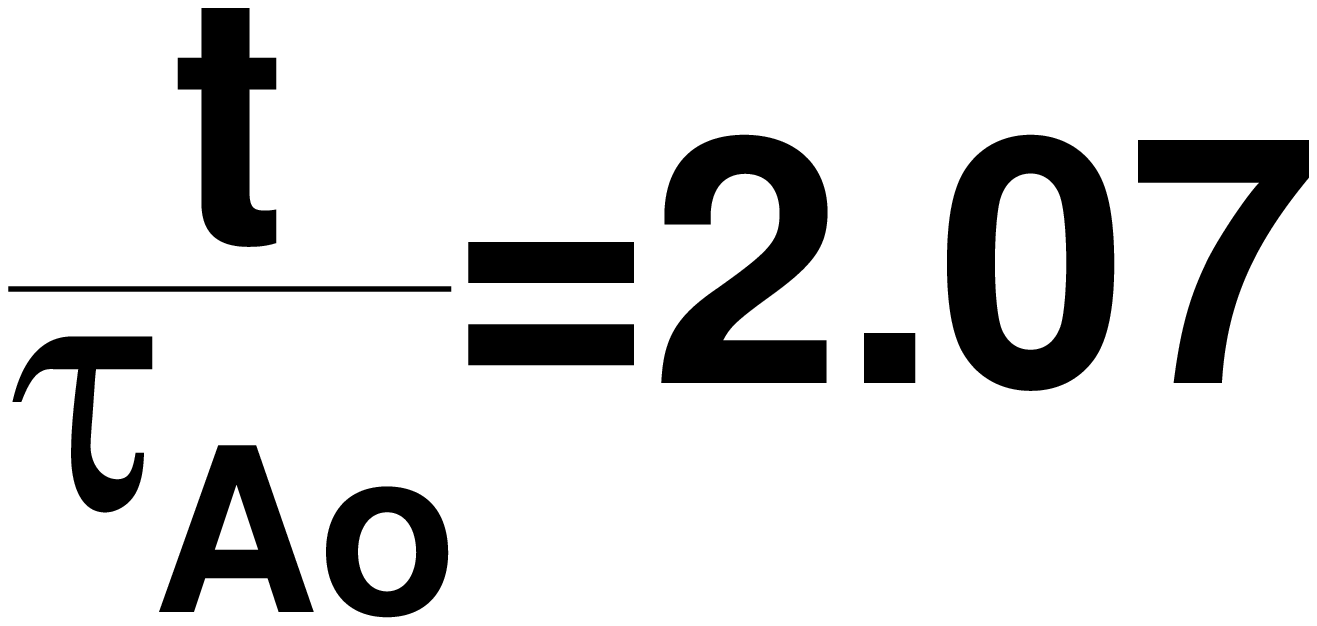}}
\subfloat{\includegraphics[height=0.225\textwidth]{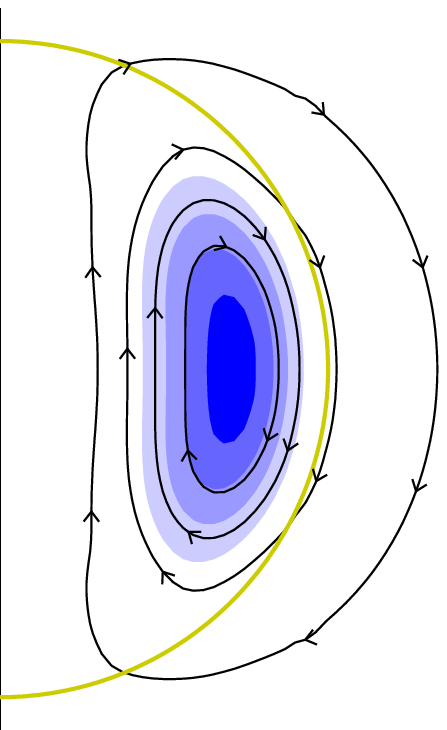}}\\
\subfloat{\includegraphics[width=0.15\textwidth,height=0.15\textwidth]{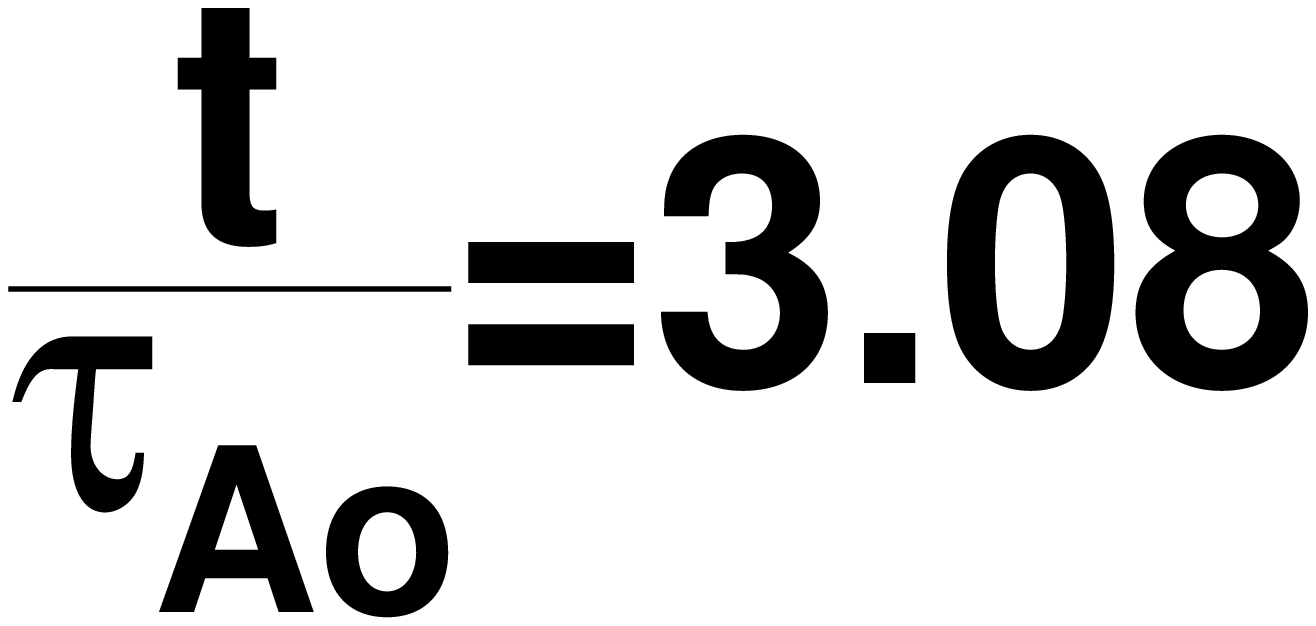}}
\subfloat{\includegraphics[height=0.225\textwidth]{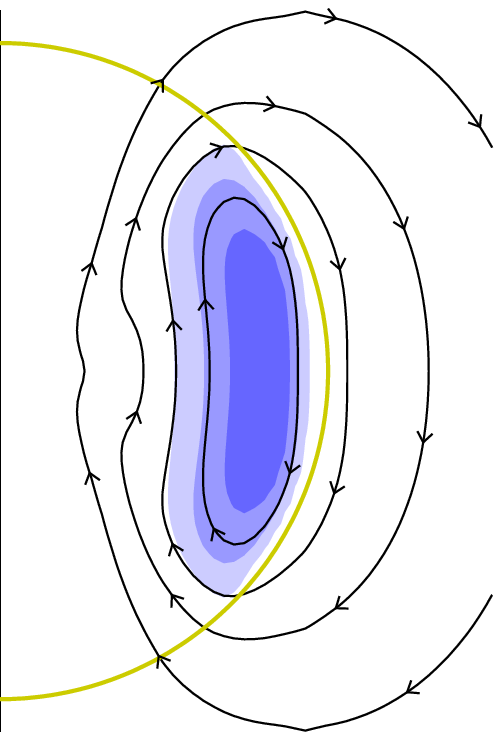}}\\
\subfloat{\includegraphics[width=0.15\textwidth,height=0.15\textwidth]{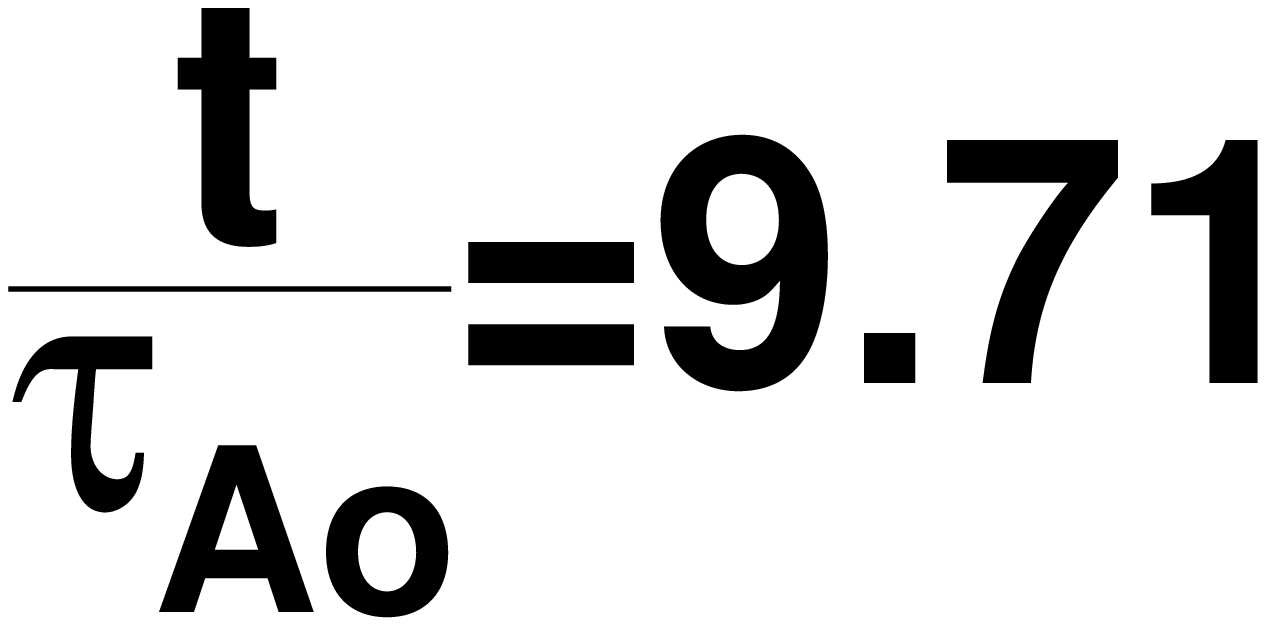}}
\subfloat{\includegraphics[height=0.225\textwidth]{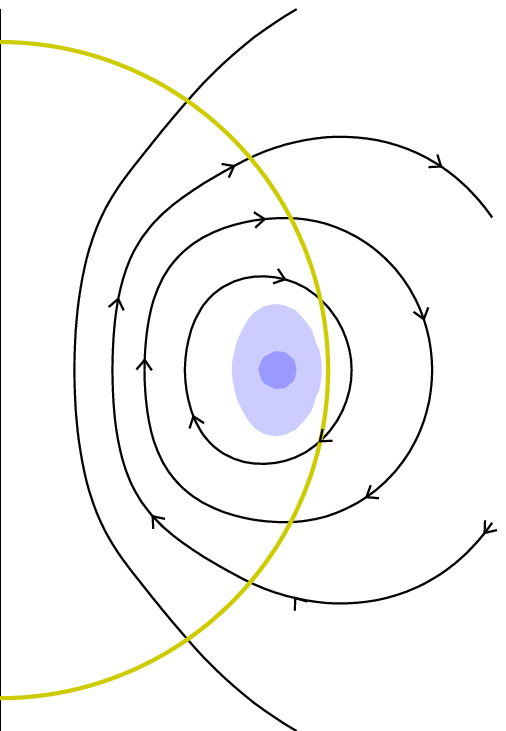}}\\
\end{tabular}
\caption{Meridional projection of the poloidal field lines and toroidal contours (drawn as $\bar{\omega}B_{\phi}$, where $\bar{\omega}$ is the cylindrical radius, and the color scale is constant for all snapshots), at times 0.28, 2.07, 3.08, and 9.71 $\tau_{\rm{Ao}}$ in a barotropic model.  At time 0.28$\tau_{\rm{Ao}}$ the configuration has not changed much from the initial state.  By time 2.07 $\tau_{\rm{Ao}}$ the torus has drifted towards the radius of the star.  After 3.08 $\tau_{\rm{Ao}}$ the toroidal field has risen into the atmosphere where it decays.  By time 9.71 $\tau_{\rm{Ao}}$ nearly all of the torus has decayed.}\label{m0plots}
\end{figure}

We have analyzed the instability of the $m$-modes in the barotropic models, with the same method as the stably stratified models (See Section \ref{n3section}).  Figure \ref{mmodes} shows the  evolution of the kinetic energies for each of the \textit{m}=0-3 modes for models with four different initial $\frac{E_{\rm pol}}{E_{\rm tot}}$ values.  In all cases, the \textit{m}=0 mode is the dominant instability, quickly rising to a value of a few hundredths of a percent of the total initial magnetic energy, at which point it remains at this range.  Even in the cases where $\frac{E_{\rm pol}}{E_{\rm tot}} > 0.8$ and $\frac{E_{\rm pol}}{E_{\rm tot}}< 0.008$, which in the stably stratified star were found to be unstable to the \textit{m}=2 and \textit{m}=1 modes respectively, the \textit{m}=0 mode is the dominant instability in the barotropic models.  To help visualize what the \textit{m}=0 mode looks like, Figure \ref{m0plots} shows that the cause of the \textit{m}=0 mode is the rise of the torus, driven by a combination of its own buoyancy and the pressure of the enclosed poloidal field.  Thus, the stable stratification seems to play a very strong role in suppressing the \textit{m}=0 mode, as it always dominates in the absence of stable stratification.

In order to see how robustly this mode dominates in de-stabilizing the magnetic equilibria, we have investigated what happens when an initial non-axially symmetric perturbation is added to the system.  To do so, we introduced perturbations to the $v_{\rm z}$ component of the velocity along the equatorial plane within the star.  %across the entire radius, up to a height of $.2R$ in $\pm$ $z$-direction.  
One case included an initial $m$=1 perturbation for a model with initial $\frac{E_{\rm pol}}{E_{\rm tot}}$=0.03, a second model contained an $m$=2 perturbation with initial   $\frac{E_{\rm pol}}{E_{\rm tot}}$=0.74, while the third model contained perturbations to $m$=1, 2, and 3 modes for an initial  $\frac{E_{\rm pol}}{E_{\rm tot}}$=0.5.  The strength of all perturbations was the same, with each mode containing an initial $\frac{E_{\rm{kin}}}{E_{\rm{mag}}}$=.01\%, roughly of the same order that the $m$=0 mode was seen to reach in the non-perturbed models.  We again found that regardless of the initial perturbations added, the $m$=0 mode always became the dominant mode of instability.

\begin{figure}
\begin{center}
\begin{tabular}{ccc}
\subfloat{\includegraphics[width=0.115\textwidth,,height=0.115\textwidth]{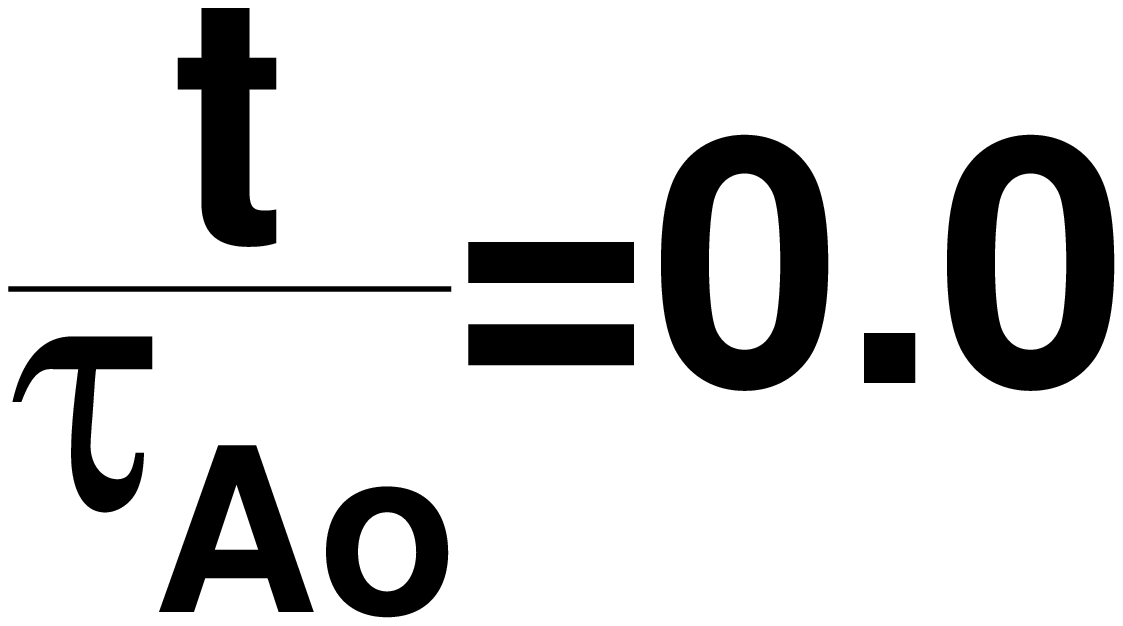}}
   & \subfloat{\includegraphics[height=0.115\textwidth]{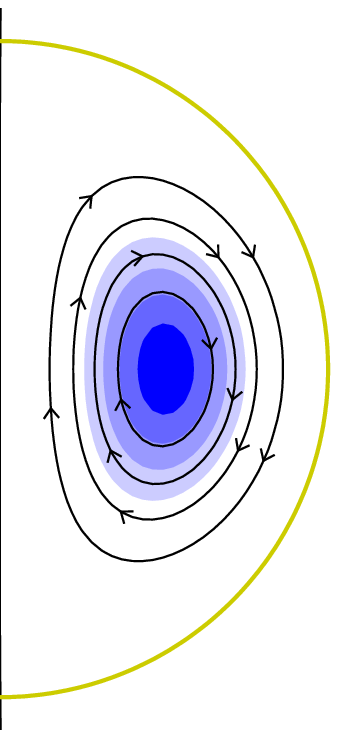}}
   & \subfloat{\includegraphics[height=0.115\textwidth]{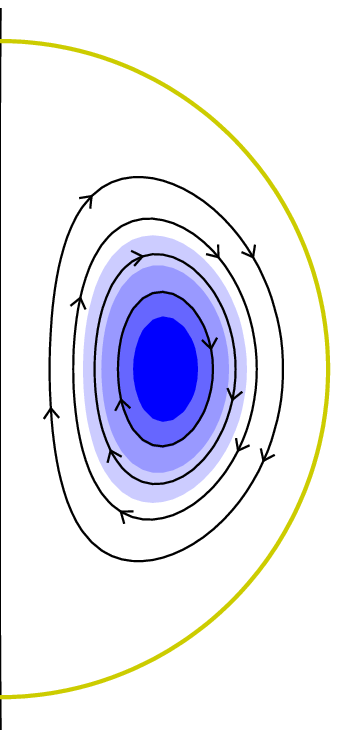}}\\
\subfloat{\includegraphics[width=0.115\textwidth,,height=0.115\textwidth]{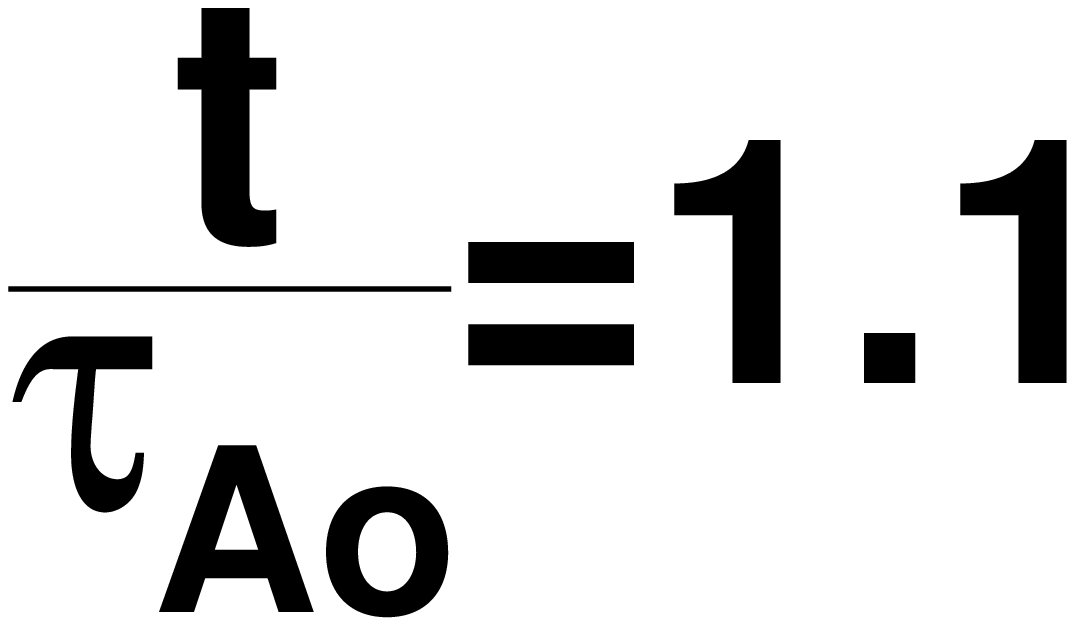}}
   &\subfloat{\includegraphics[height=0.115\textwidth]{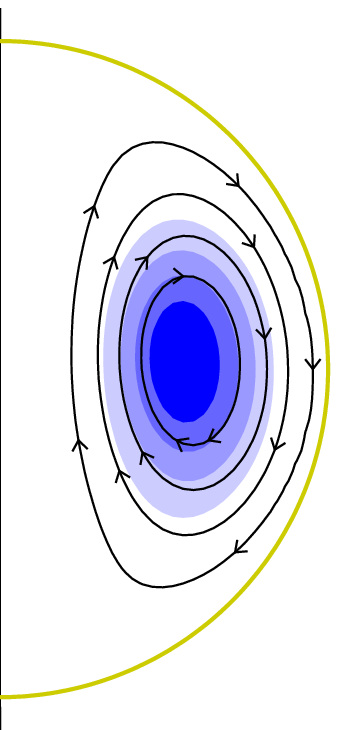}}
   & \subfloat{\includegraphics[height=0.115\textwidth]{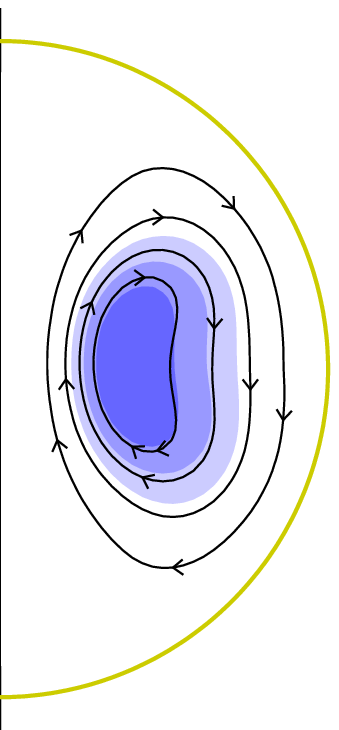}}\\
\subfloat{\includegraphics[width=0.115\textwidth,,height=0.115\textwidth]{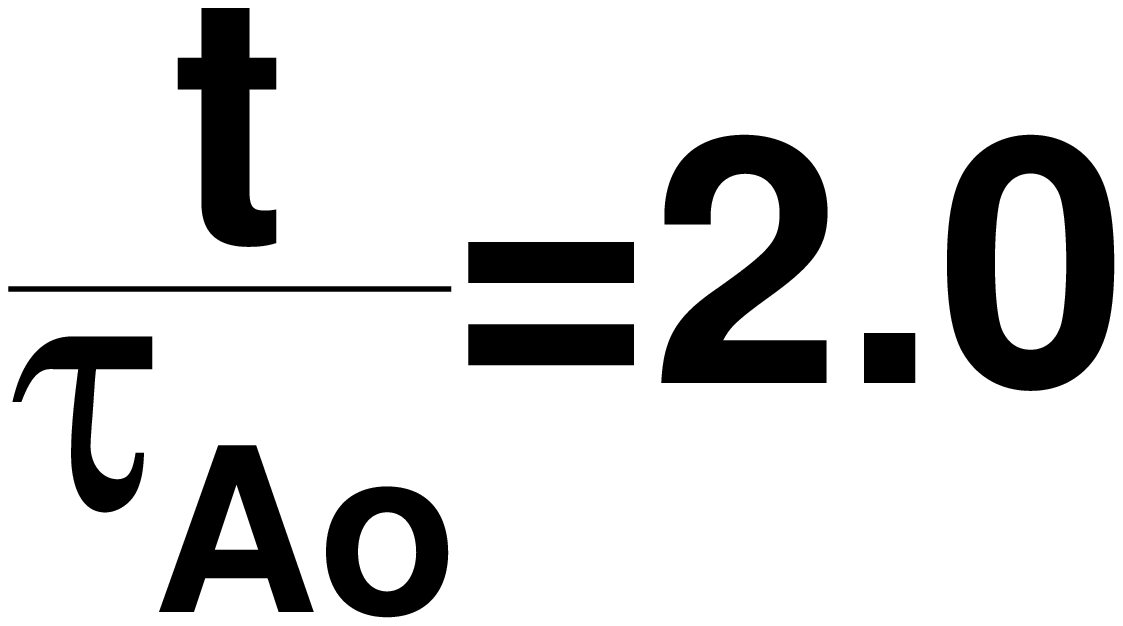}}
   & \subfloat{\includegraphics[height=0.115\textwidth]{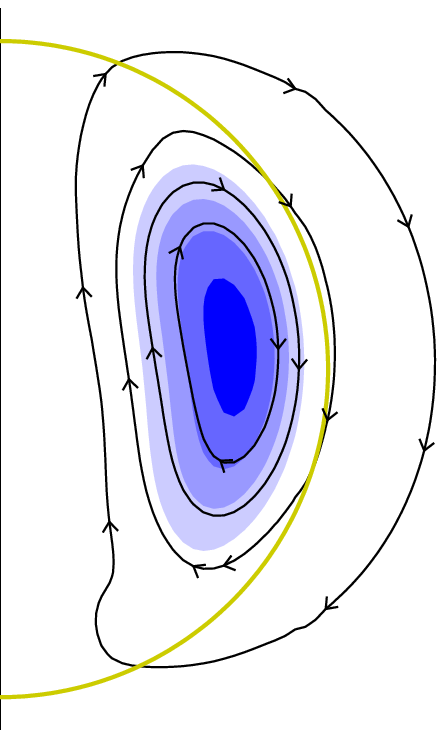}}
   & \subfloat{\includegraphics[height=0.115\textwidth]{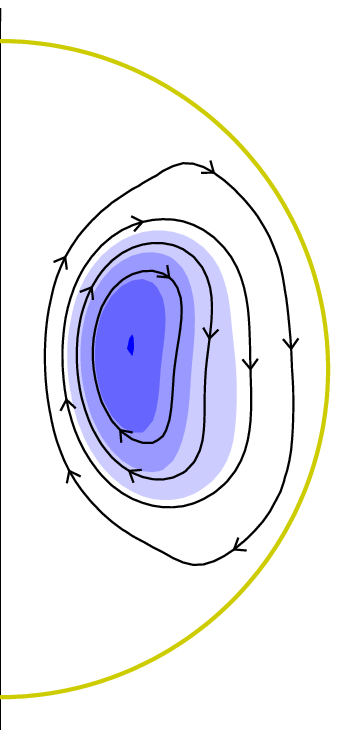}}\\
\subfloat{\includegraphics[width=0.115\textwidth,,height=0.115\textwidth]{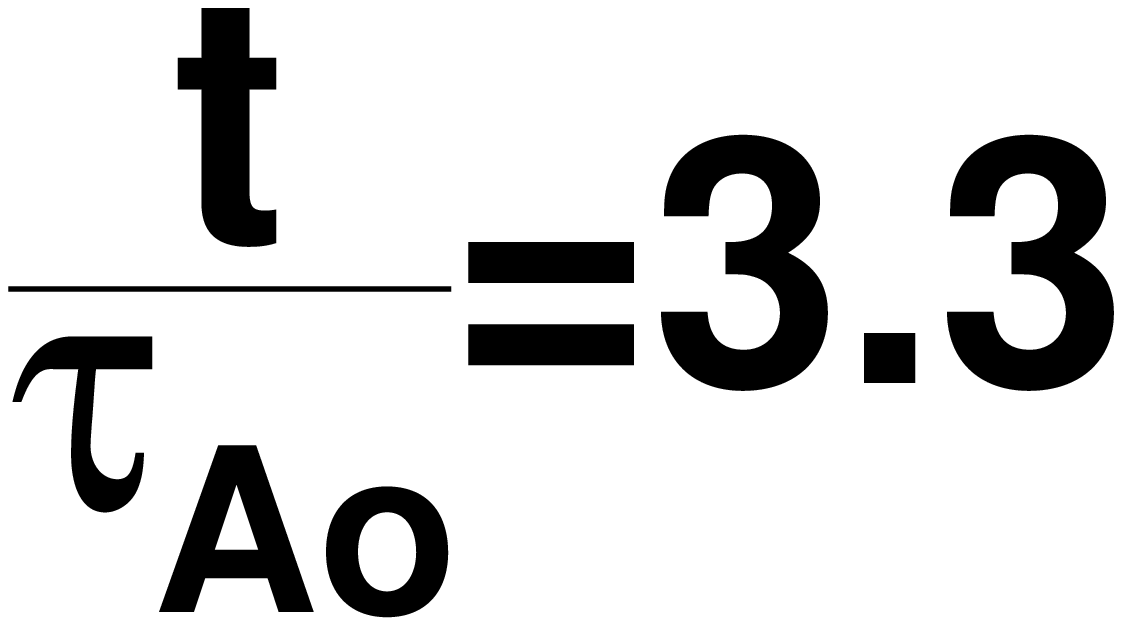}}
   & \subfloat{\includegraphics[height=0.115\textwidth]{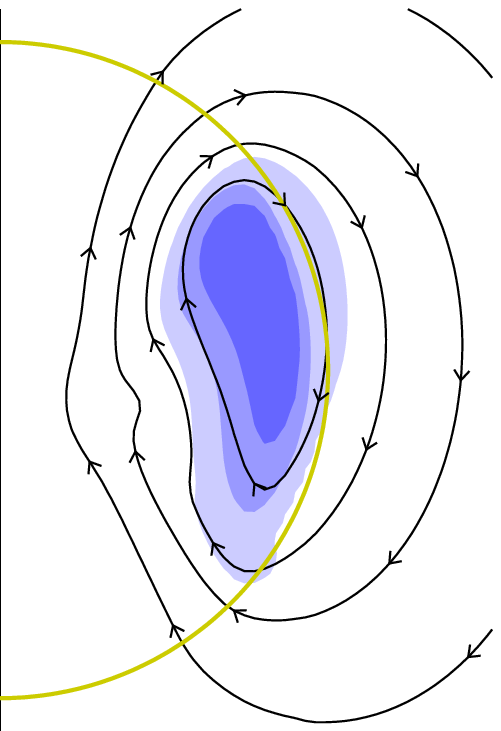}}
   & \subfloat{\includegraphics[height=0.115\textwidth]{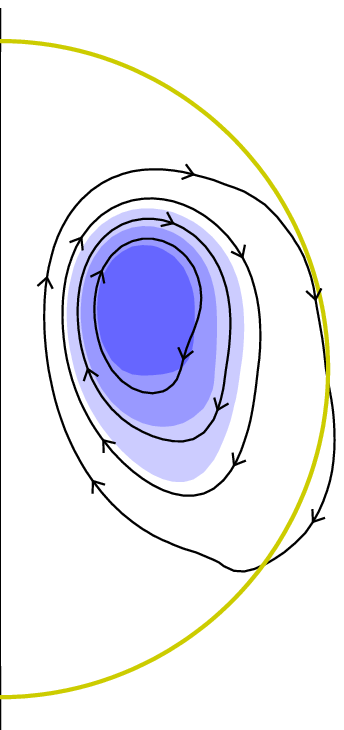}}\\
\subfloat{\includegraphics[width=0.115\textwidth,,height=0.115\textwidth]{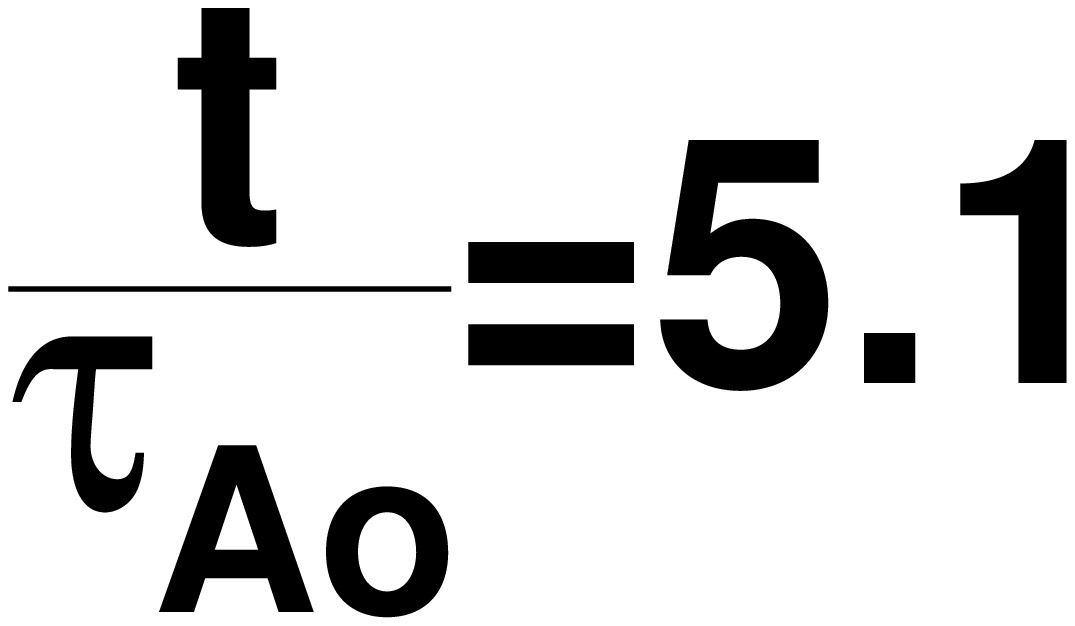}}
   & \subfloat{\includegraphics[height=0.115\textwidth]{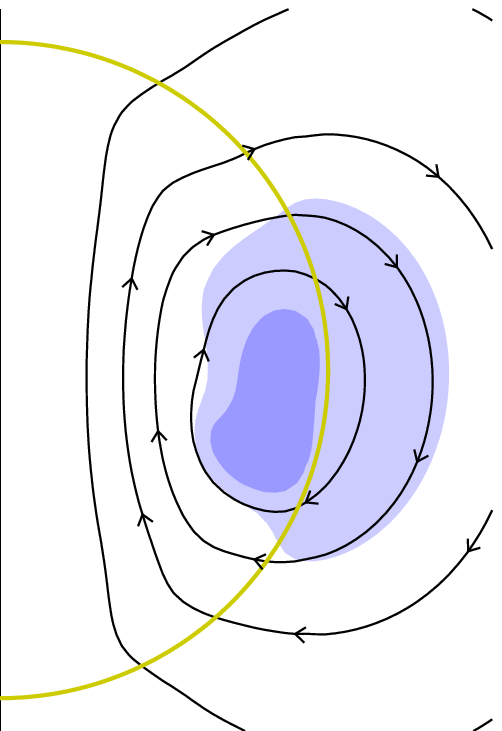}}
   & \subfloat{\includegraphics[height=0.115\textwidth]{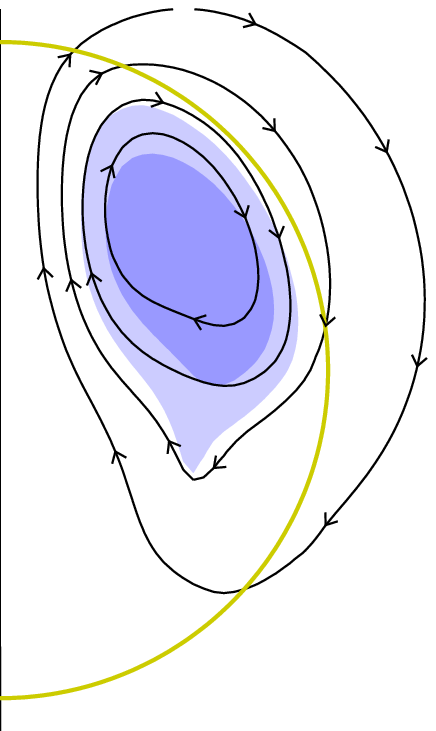}}\\
\subfloat{\includegraphics[width=0.115\textwidth,,height=0.115\textwidth]{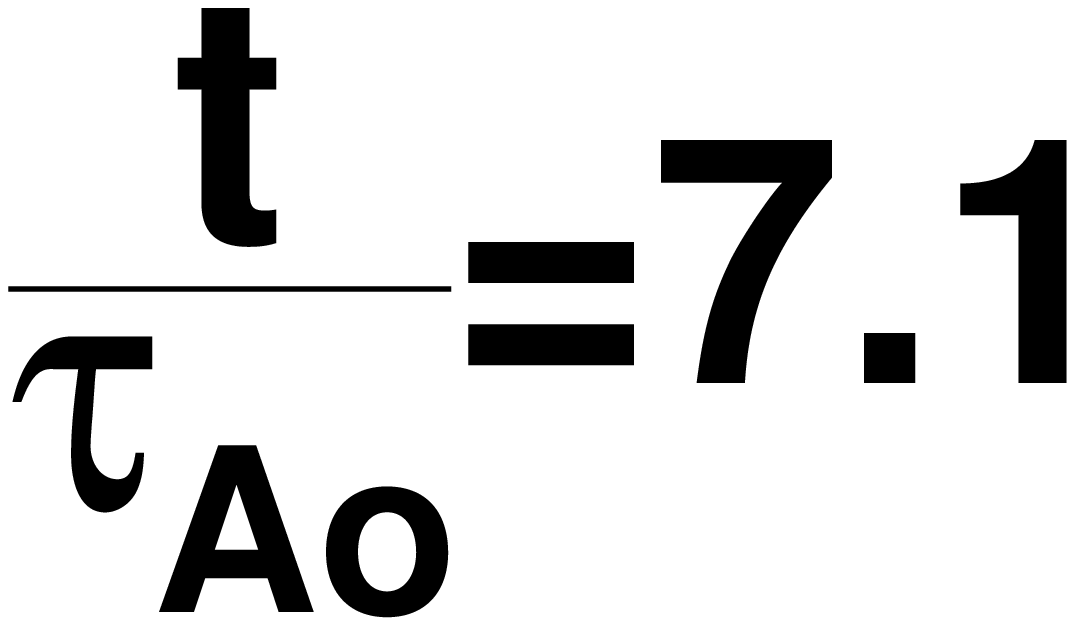}}
   & \subfloat{\includegraphics[height=0.115\textwidth]{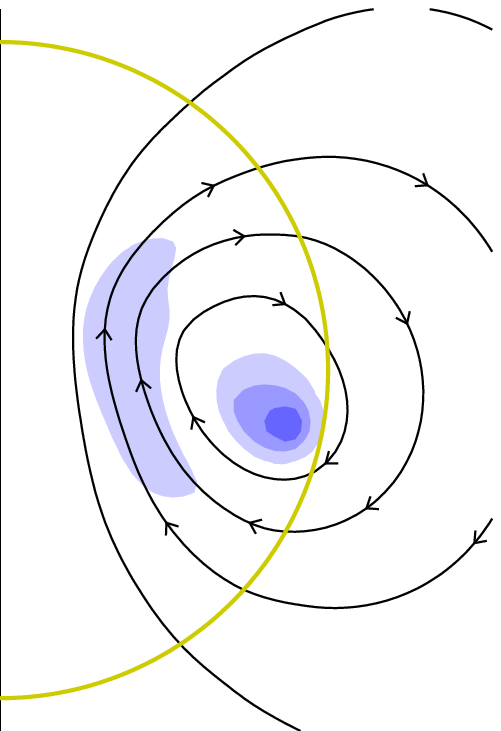}}
   & \subfloat{\includegraphics[height=0.115\textwidth]{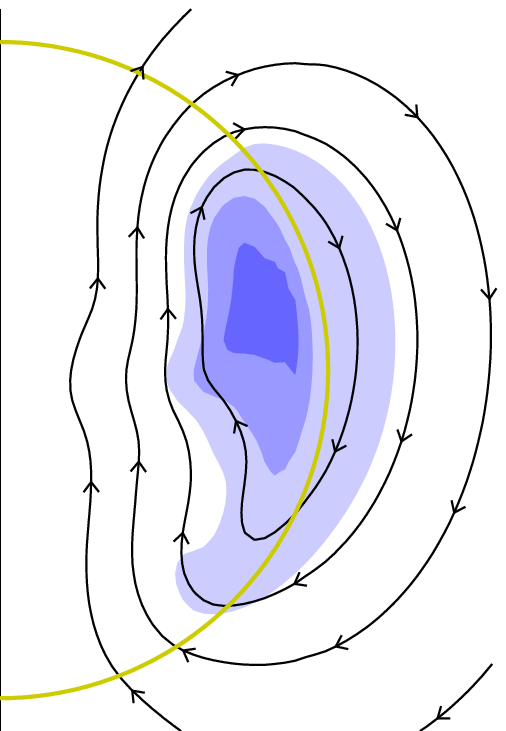}}\\
\end{tabular}
\end{center}
\caption{Snapshots of the poloidal field (black lines) and toroidal field (drawn as $\bar{\omega}B_{\phi}$) (blue color scale), for magnetic field configurations with an initial $\frac{E_{\rm pol}}{E_{\rm tot}}$ of 0.5 (left panels), and 0.24 (right panels), which were perturbed axially symmetrically, by moving the configuration .016$R$ above the equator, causing a North-South asymmetry in barotropic models.  The snapshots are at time: 0., 1.1, 2.0, 3.3, 5.1, and 7.1 $\tau_{\rm{Ao}}$, from top to bottom.  By time 1.1 $\tau_{\rm{Ao}}$, it is possible to see that the toroidal field in the 0.5 model has risen to slightly larger radii, while the 0.24 model's torus has contracted to smaller radii.  By time 2.0 $\tau_{\rm{Ao}}$, the tori in both models begin to move towards the northern pole, whilst continuing their previously seen radial movement.  By time 3.3 $\tau_{\rm{Ao}}$, the tori in each model have moved noticeably towards higher latitudes, and the magnetic field in the 0.24 model begins to rise to larger radii.  By 5.1 $\tau_{\rm{Ao}}$, the torus of the 0.24 model has continued its northerly path, while the torus in the 0.5 model has begun to decay in the atmosphere.  (Note there is some toroidal component outside the star, this is because the enhanced diffusion utilized in the atmosphere takes a bit of time to make the field relax to a potential field.)  By the final time step of 7.1 $\tau_{\rm{Ao}}$, the torus of the 0.5 model has decayed significantly, while the torus of the 0.24 model has started to reach the atmosphere where it is now experiencing a fast decay of the torus.  In both cases, the magnetic field configurations remain axisymmetric throughout this process.  }
\label{NSperts}
\end{figure}

In addition to non-axially symmetric perturbations, we have run two simulations with a non-equatorially symmetric perturbation.  The perturbation was created by offsetting the magnetic field configuration by 0.016$R$ in the Northern direction, for two initial $\frac{E_{\rm pol}}{E_{\rm tot}}$ values of 0.5 and 0.24.  Snapshots of the time evolution of the poloidal and toroidal field lines can be seen in Figure \ref{NSperts}.  The evolution in these northerly perturbed models is similar to their non-perturbed counterparts, with the only difference being that these models have a slight tendency to evolve away from the equator.

\section{Conclusions}\label{concsec}

We have conducted two different kinds of numerical experiments.

In the first kind, we started with disordered initial fields. In the previously studied case of
stably stratified stellar models, we confirmed that the magnetic field evolved
into an ordered, stable, roughly axially symmetric configuration with
comparable poloidal and toroidal fields. In the barotropic cases we
studied, this never happened, and the magnetic energy decayed much
faster than expected from diffusion.

In the second kind, we started with a smooth, axially symmetric magnetic field
with poloidal and toroidal components. We confirmed that, in the stably stratified
star, there are stable hydromagnetic equilibria for a fairly wide range of
values of the initial fraction of poloidal to total magnetic energy, $0.008 < \frac{E_{\rm pol}}{E_{\rm tot}} < 0.8$.
Outside this range, the field decayed through non-axisymmetric modes, $m=1$
in the toroidally dominated and $m=2$ in the poloidally dominated case.
In contrast, in the barotropic case, we found no stable configurations
(exploring a fairly large parameter space), and the field always decayed through an axially
symmetric ($m=0$) instability in which the toroidal flux rose radially
out of the star, being dissipated in the atmosphere.

Although far from a rigorous mathematical proof, this provides
strong evidence (added to that previously provided by
\citealt{Lander_2012}) that there are no stable equilibria in
barotropic stars. It also strongly supports the intuition that, in
a stably stratified star, the buoyancy force strongly constrains
the potential instabilities by impeding any substantial radial
motions, whereas such motions are not hindered in the barotropic
case, and in fact these motions destabilize the magnetic
equilibrium.

As argued above, the stabilization provided to the magnetic field
by the buoyancy force can be roughly quantified by comparing the
Brunt-V\"ais\"al\"a frequency $N$ to the Alfv\'en frequency
$\omega_A$. It will only be effective if $N\gtrsim \omega_A$,
whereas in the opposite case the star would behave as if it were
barotropic, not being able to contain a stable magnetic field. If
there are dissipative processes effectively eroding the stable
stratification on long time scales, they will lead to a
destabilization and eventual decay of the magnetic field, unless
it can be stabilized, e.~g., by the solid crust of a neutron star.
In fact, in the neutron star case, this effect has been argued to
act on astrophysically relevant time scales, which does not seem
to be the case in white dwarfs and upper main sequence stars
\citep{Reisenegger_2009}, with the possible exception of very
massive O stars, which are radiation-dominated and thus only
weakly stratified.  

Note that the numerical models used here, which assume a chemically uniform, classical ideal gas, stably stratified by an entropy gradient, clearly do not directly apply to degenerate stars, such as white dwarfs and neutron stars. However, we have no reason to doubt that the essential physics, in particular the competition between magnetic and buoyancy forces, is the same in these cases, and the (very rough) condition $N \gtrsim \omega_{A}$ is still required to stabilize a hydromagnetic equilibrium in these cases, even in the neutron star case, where $N$ is due to a composition gradient. Additional changes in the neutron star case are the presence of a stabilizing solid crust, as well as superconducting and superfluid regions in the interior, which modify the form of the Lorentz force and the dynamical equations and thus do not allow a direct application of the present results, although some of their features might carry over to this regime.

\section{Acknowledgements}
This work was supported by the DFG-CONICYT International Collaboration 
Grant DFG-06, 
%Check whether there is a different number for the German part, and  whether there
%are other German sources to be mentioned!!!
FONDECYT Regular Project 1110213, and Proyecto de Financiamiento Basal
PFB-06/2007.  Some of the figures were made with VAPOR (www.vapor.ucar.edu).

\begin{appendix}
\section{Numerical acceleration scheme}
The scaling of magnetic fields that evolve from different initial amplitudes (but otherwise identical configuration) with time is described in Section \ref{times}. The practical implementation in the MHD code is as follows. We need to make a distinction between the field ${\bf B}_{\rm num}$ and time $t_{\rm num}$ in the accelerated code, and the physical field  ${\bf B}$ and time $t$ reconstructed from it.
The MHD induction equation 
\begin{equation}
{\partial{\bf B_{\rm num}}\over\partial t_{\rm num}}=-{1\over c}\nabla\times{\bf E},
\end{equation}
is first evolved over a time step $\Delta t_{\rm num}$, where ${\bf E}=-{{\bf v\times B_{\rm num}}\over c}$, to yield an intermediate update $\Delta{\bf B}^\prime$. The magnetic energy over the numerical volume $V$ is measured from the result of this time step:
\begin{equation}
E_{\rm num}=\int B_{\rm num}^2/8\pi\,{\rm d}V,
\end{equation}
and the evolution time scale $\tau_{\rm num}$ (in the numerical time unit) is calculated:
\begin{equation}
\tau_{\rm num}=4\pi E_{\rm num}\,/\int{\bf B}_{\rm num}\cdot \partial{\bf B}_{\rm num}/\partial t_{\rm num}.
\end{equation}
The intermediate update $\Delta{\bf B}^\prime$ is modified:
\begin{equation}
\Delta{\bf B}=\Delta{\bf B}^\prime+{\bf B}^\prime\Delta t_{\rm num}/2\tau_{\rm num},
\end{equation}
which brings the numerical Alfv\'en crossing time back to its initial value $\tau_{\rm{Ao}}$.
The new value of the physical magnetic energy $E$ has changed by the 
amount 
\begin{equation}
\Delta E=-E\Delta t_{\rm num}/\tau_{\rm num},
\end{equation}
and the physical time coordinate by the amount
\begin{equation}
\Delta t=\Delta t_{\rm num}\sqrt{E_{\rm num}/E}.
\end{equation}
\end{appendix}

\bibliography{paper.bib}
\end{document}